\PassOptionsToPackage{unicode}{hyperref}
\PassOptionsToPackage{hyphens}{url}
\PassOptionsToPackage{dvipsnames,svgnames,x11names}{xcolor}
\documentclass[
  10pt,
]{article}
\usepackage{xcolor}
\usepackage[margin=0.9in,top=0.8in,bottom=0.8in]{geometry}
\usepackage{amsmath,amssymb}
\usepackage{iftex}
\ifPDFTeX
  \usepackage[T1]{fontenc}
  \usepackage[utf8]{inputenc}
  \usepackage{textcomp} 
\else 
  \usepackage{unicode-math} 
  \defaultfontfeatures{Scale=MatchLowercase}
  \defaultfontfeatures[\rmfamily]{Ligatures=TeX,Scale=1}
\fi
\usepackage{lmodern}
\ifPDFTeX\else
\fi
\IfFileExists{upquote.sty}{\usepackage{upquote}}{}
\IfFileExists{microtype.sty}{
  \usepackage[]{microtype}
  \UseMicrotypeSet[protrusion]{basicmath} 
}{}
\makeatletter
\@ifundefined{KOMAClassName}{
  \IfFileExists{parskip.sty}{%
    \usepackage{parskip}
  }{
    \setlength{\parindent}{0pt}
    \setlength{\parskip}{6pt plus 2pt minus 1pt}}
}{
  \KOMAoptions{parskip=half}}
\makeatother
\usepackage{longtable,booktabs,array}
\usepackage{calc} 
\usepackage{etoolbox}
\makeatletter
\patchcmd\longtable{\par}{\if@noskipsec\mbox{}\fi\par}{}{}
\makeatother
\IfFileExists{footnotehyper.sty}{\usepackage{footnotehyper}}{\usepackage{footnote}}
\makesavenoteenv{longtable}
\usepackage{graphicx}
\makeatletter
\newsavebox\pandoc@box
\newcommand*\pandocbounded[1]{
  \sbox\pandoc@box{#1}%
  \Gscale@div\@tempa{\textheight}{\dimexpr\ht\pandoc@box+\dp\pandoc@box\relax}%
  \Gscale@div\@tempb{\linewidth}{\wd\pandoc@box}%
  \ifdim\@tempb\p@<\@tempa\p@\let\@tempa\@tempb\fi
  \ifdim\@tempa\p@<\p@\scalebox{\@tempa}{\usebox\pandoc@box}%
  \else\usebox{\pandoc@box}%
  \fi%
}
\def\fps@figure{htbp}
\makeatother
\NewDocumentCommand\citeproctext{}{}
\NewDocumentCommand\citeproc{mm}{%
  \begingroup\def\citeproctext{#2}\cite{#1}\endgroup}
\makeatletter
 \let\@cite@ofmt\@firstofone
 \def\@biblabel#1{}
 \def\@cite#1#2{{#1\if@tempswa , #2\fi}}
\makeatother
\newlength{\cslhangindent}
\newlength{\csllabelwidth}
\newenvironment{CSLReferences}[2] 
 {\begin{list}{}{%
  \setlength{\itemindent}{0pt}
  \setlength{\leftmargin}{0pt}
  \setlength{\parsep}{0pt}
  \ifodd #1
   \setlength{\leftmargin}{\cslhangindent}
   \setlength{\itemindent}{-1\cslhangindent}
  \fi
  \setlength{\itemsep}{#2\baselineskip}}}
 {\end{list}}
\usepackage{calc}

\providecommand{\tightlist}{%
  \setlength{\itemsep}{0pt}\setlength{\parskip}{0pt}}
\usepackage[T1]{fontenc}
\usepackage{charter}
\usepackage[scaled=0.9]{helvet}

\usepackage{textcomp}
\DeclareUnicodeCharacter{03C7}{\ensuremath{\chi}}
\DeclareUnicodeCharacter{03C4}{\ensuremath{\tau}}
\DeclareUnicodeCharacter{00B2}{\textsuperscript{2}}
\DeclareUnicodeCharacter{2212}{\ensuremath{-}}
\usepackage[table]{xcolor}
\definecolor{ink}{HTML}{1B2430}
\definecolor{ink2}{HTML}{465061}
\definecolor{muted}{HTML}{6E7886}
\definecolor{rule}{HTML}{C9D1DA}
\definecolor{shadecolor}{HTML}{F1F4F8}
\definecolor{accent}{HTML}{1F4E8C}
\definecolor{highlight}{HTML}{2A78D6}
\color{ink}
\usepackage{booktabs}
\usepackage{float}
\usepackage{needspace}   
\usepackage{framed}
\usepackage[nobottomtitles*]{titlesec}   
\titleformat{\section}{\sffamily\bfseries\large\color{ink}}{\thesection}{0.6em}{}[{\color{rule}\titlerule}]
\titleformat{\subsection}{\sffamily\bfseries\normalsize\color{ink}}{\thesubsection}{0.6em}{}
\titlespacing*{\subsection}{0pt}{8pt plus 2pt minus 2pt}{3pt}
\titlespacing*{\section}{0pt}{11pt plus 2pt minus 2pt}{5pt}
\usepackage[font={small,color=ink2},labelfont={sf,bf},labelsep=period,skip=4pt]{caption}
\usepackage{fancyhdr}
\usepackage{enumitem}
\setlist{nosep, leftmargin=1.4em, itemsep=2pt}

\usepackage{etoolbox}
\AtBeginEnvironment{longtable}{\small\sffamily}
\makeatletter
\renewcommand{\maketitle}{%
  {\bfseries\Large\raggedright\@title\par}\vspace{5pt}
  {\small\ifx\@author\@empty\else
    {\def\and{\end{tabular}\hspace{2.5em}\begin{tabular}[t]{@{}l@{}}}%
     \color{ink}\noindent\begin{tabular}[t]{@{}l@{}}\@author\end{tabular}\par}\vspace{4pt}\fi
   {\color{ink2}\@date}\par}
  \vspace{8pt}{\color{rule}\hrule height 0.6pt}\vspace{10pt}}
\makeatother
\renewenvironment{abstract}{\small\noindent{\sffamily\bfseries Abstract.}\ }{\par\vspace{6pt}}
\usepackage{bookmark}
\IfFileExists{xurl.sty}{\usepackage{xurl}}{} 
\hypersetup{
  pdftitle={Identical Runs{,} Different Results: Benchmarking AI Coding Agents on Open-Weight Models},
  pdfauthor={Eduardo Ariño de la Rubia Central European University rubiae@ceu.edu; Szilard Pafka Epoch szilard@epoch.com},
  colorlinks=true,
  linkcolor={ink2},
  filecolor={Maroon},
  citecolor={ink2},
  urlcolor={ink2},
  pdfcreator={LaTeX via pandoc}}

\title{Identical Runs, Different Results: Benchmarking AI Coding Agents
on Open-Weight Models}
\author{Eduardo Ariño de la Rubia\\
Central European University\\
\href{mailto:rubiae@ceu.edu}{\nolinkurl{rubiae@ceu.edu}} \and Szilard
Pafka\\
Epoch\\
\href{mailto:szilard@epoch.com}{\nolinkurl{szilard@epoch.com}}}
\date{September 2026}

\begin{document}
\maketitle
\begin{abstract}
Repeated runs of the same coding agent are known to give different
benchmark scores. We ask what that variation means for a team running an
agent on its own task, by intensive replication on one machine-learning
task: an agent improves the training code of an XGBoost classifier for
airline delays, and a holdout it never sees scores the result. Across
584 runs, we compare six agents on six open-weight model endpoints, run
six agent-model pairings 52 times each under fixed settings, and repeat
three of them on a larger model from the same family. Identical runs of
one pairing varied more than the pairings differed from one another, so
comparisons of a few runs ranked them unreliably; resolving the agent
differences we observed would take tens to more than a hundred runs of
each. Runs on the larger model scored clearly higher, but by less than
one run-to-run standard deviation, and the gap was more than twice as
large with one agent as with the others. Fewer than one run in twenty
broke the task's data rules, but those runs held the highest scores.
Rejecting those runs first and keeping the best compliant result among a
few attempts reliably improved the delivered model, even though a few
runs could not rank the agents. On flights from a later year, the
delivered models kept only a third of their gain over the starting code.
At list prices, cost differed more than twentyfold between two agents on
the same model, mostly through the prompt cache. Agents and models
should be evaluated as pairings, over repeated attempts, with compliance
reported beside quality. Data, code and every delivered program:
https://github.com/earino/identical-runs-different-results
\end{abstract}

\section{Introduction}\label{introduction}

A coding agent is a language model inside a harness: the program that
decides what the model sees, which tools it can call and when the work
continues. Teams that adopt one face three questions. Which agent? Which
model? And how many attempts before trusting the result? Many
coding-agent evaluations report one or a few attempts per task and
emphasise averages across many tasks. A recent study of harness design
runs each of its settings once per task
(\citeproc{ref-fan2026harness}{Fan et al. 2026}); HarnessTax runs three
attempts per task on 30-task samples and averages them
(\citeproc{ref-pan2026harnesstax}{Pan et al. 2026}); MLE-bench's
guidance asks for at least three seeds because agents are high-variance
(\citeproc{ref-chan2024mlebench}{Chan et al. 2025}). Bjarnason et al.
(\citeproc{ref-bjarnason2026randomness}{2026}) have already shown that
this is not enough: ten independent runs of six agent configurations at
two temperatures on SWE-bench Verified, 60,000 trajectories in all,
moved single-run pass@1 by 2.2 to 6.0 percentage points depending on
which run was used, with standard deviations above 1.5 points even at
temperature 0. They recommend estimating pass@1 from several runs per
task and choosing the number of runs by power analysis. We take that
lesson as given. Their unit is still a benchmark score, a pass rate over
many tasks with each attempt a pass or a fail. A team running an agent
on its own task meets something else: the spread of a continuous measure
of what one task's attempts deliver, and questions a pass rate does not
raise. Did the delivered artifact break the task's rules? Is keeping the
best of several attempts worth it? Does the gain hold on data from
later?

We study that question by intensive replication on one machine-learning
task, because such a task makes an agent's work scorable without a human
grader. The agent is handed a working XGBoost model that predicts
whether a US airline flight will leave at least 15 minutes late, from
eight fields such as carrier, route and departure time, and it edits the
training code to make the model predict better. A holdout of one million
later flights, which no agent sees, scores the delivered model by AUC:
the chance that the model ranks a late flight above an on-time one. The
score is continuous and has room to rise, each run is cheap to repeat,
and breaking the task's data rules leaves traces in the delivered code.

Three studies use this task. The first is broad: six agents on six
open-weight model endpoints, three runs each. The second is deep: three
agents on two models, each pairing run 52 times with the data, prompt,
budget, machine type and parallelism fixed, 312 runs in all. The third
makes a planned change of model: the same three agents on a larger model
from the same family, another 156 runs.

Run-to-run variation, harness effects that depend on the model,
best-of-k evaluation and rule-breaking by agents are each established
(Section 2); we do not claim to have found them. What we add are
measurements a team can act on, for one task under one protocol: the
distribution of artifact quality, observed compliance and the value of
repeat-and-select policies, measured together. The central distinction
is between selecting a useful artifact and ranking configurations. A few
attempts, with rule-breaking attempts rejected before the best is
chosen, reliably buy a better artifact; telling configurations apart
takes tens of runs of each; and scores on a later year shrink every
gain. Intensive replication also lets us set the variation within a
pairing against a planned model substitution and against exploratory
agent-by-model interactions. A separate analysis examines how the
deployed pairing affects token use and caching. We report six results.

\begin{enumerate}
\def\labelenumi{\arabic{enumi}.}
\tightlist
\item
  \textbf{One run is a draw.} The median pairing's run-to-run standard
  deviation over compliant runs, 0.0107 AUC, is larger than the 0.0095
  spread between the six pairing means. Three-run comparisons put the
  weaker pairing ahead 28 to 44 percent of the time, and for the agent
  differences observed here the usual approximation calls for roughly 39
  to 113 runs per arm.
\item
  \textbf{Compliance failures sit at the top of the ranking.} Ten of 312
  runs trained on the labelled evaluation file or computed features from
  the batch they were scoring, and they include the seven highest
  scores. Excluding them lowers the best score from 0.8293 to 0.7695 and
  leaves the spread between pairing means almost unchanged.
\item
  \textbf{A few attempts buy a better artifact, not a reliable ranking.}
  The best compliant artifact among three attempts, chosen on the
  evaluation set, has a median holdout AUC 0.0081 above one attempt (95
  percent interval 0.0063 to 0.0098), and three attempts returned a
  compliant artifact with an estimated probability of at least 99.88
  percent.
\item
  \textbf{A planned model upgrade is about one run of noise, and its
  size depends on the agent.} Runs on the larger model scored 0.0091 AUC
  higher, 7.1 standard errors from zero, yet one run of each still
  favoured the smaller model 28 percent of the time. pi's gain, 0.0152,
  exceeded the other two agents' by 2.7 and 3.3 standard errors.
\item
  \textbf{Cost depends on the pairing, largely through the prompt
  cache.} In a rate-card projection for one observed comparison, the
  same job cost \$0.08 a run through pi and \$1.86 through Claude Code
  on one endpoint, and most of the gap follows the share of input the
  endpoint served from its cache.
\item
  \textbf{Much of the gain belongs to the year the agents tuned on.}
  Scored on one million flights from 2007, the same artifacts kept a
  third of their gain over the starting code, the spread between runs
  halved, and the larger model's advantage fell to 0.44 run-to-run
  standard deviations. Every finding kept its direction, but pi's larger
  gain no longer separated from the other agents'.
\end{enumerate}

Section 2 places these results in prior work. Section 3 describes the
task, the agents and the three designs. Section 4 reports the results by
finding rather than by study, and Section 5 discusses what they mean for
evaluating agents, what a difference in AUC is worth, and the
limitations.

\section{Related work}\label{related-work}

\subsection{Harnesses, models and their
cost}\label{harnesses-models-and-their-cost}

Agent evaluation has moved from treating the model as the evaluated
object to studying the system around it. SWE-agent showed that an
interface designed for the model makes a software-engineering agent more
capable than a plain shell does (\citeproc{ref-yang2024sweagent}{Yang et
al. 2024}). Lewis (\citeproc{ref-lewis2026same}{2026}) held a model and
its tasks fixed and changed only how the harness managed context. On 169
SWE-bench Verified tasks under a tight context window, complete
solutions rose from 43 to 72, and the change transferred to three other
models; the paper concludes that ``coding-agent evaluations should treat
the model and harness together as the tested solver''. Fan et al.
(\citeproc{ref-fan2026harness}{2026}) ablate the components of one
harness across four open-weight models, running each setting once per
task and comparing settings with paired tests over tasks, and find
effects that depend on the model: planning is ``an accuracy scaffold for
weaker models'' and ``a cost saver for stronger models''. HarnessTax
(\citeproc{ref-pan2026harnesstax}{Pan et al. 2026}) pairs seven models
with Claude Code, Codex CLI and pi on 30-task samples of SWE-bench Lite
and Terminal-Bench 2.0 (\citeproc{ref-jimenez2024swebench}{Jimenez et
al. 2024}; \citeproc{ref-merrill2026terminalbench}{Merrill et al.
2026}), averages three attempts per task, and finds that the harness
moves cost far more than success; its strongest pairing solved 97.8
percent of its attempts on the sampled SWE-bench Lite tasks. HarnessTax
emphasises mean cost and success across tasks; our analysis emphasises
the distribution of artifact quality across repeated attempts on a fixed
task.

On cost, Kapoor et al. (\citeproc{ref-kapoor2024agents}{2025}) argue
that agents should be compared on cost and accuracy together. Lawrence
(\citeproc{ref-lawrence2026harness}{2026}) holds the model fixed and
routes twelve harness configurations through one gateway. Cache-aware
accounting reverses several of its cost rankings, and it shows that the
share of input read from cache is a property of the whole path from
harness through gateway to provider: the one harness speaking the
gateway's Anthropic-style dialect, Claude Code, received almost no cache
reads where the others received 56 to 85 percent. Our cache observation
(Section 4.9) is an additional operational case of the same kind, on a
different gateway, with a pattern that a dialect translation failing
everywhere would not produce.

\subsection{Agents doing machine-learning engineering, and breaking the
rules}\label{agents-doing-machine-learning-engineering-and-breaking-the-rules}

MLE-bench evaluates agents on 75 Kaggle competitions
(\citeproc{ref-chan2024mlebench}{Chan et al. 2025}). RE-Bench compares
agents with human experts on seven open-ended ML research-engineering
environments and aggregates repeated attempts as best-of-k under
different time budgets (\citeproc{ref-wijk2025rebench}{Wijk et al.
2025}). DeltaML-Bench asks agents to improve published baselines in 48
research repositories (\citeproc{ref-moukpe2026deltaml}{Moukpe et al.
2026}). These benchmarks cover far more tasks than ours; we study one
task deeply enough to estimate the distribution of repeated identical
attempts. Closest to our setting, Ferreira et al.
(\citeproc{ref-ferreira2026autoresearch}{2026}) use autoresearch, in
which an agent tunes a model by editing its training code under a fixed
compute budget, to compare agents with classical hyperparameter
optimisation: editing code narrowed the gap to classical methods without
closing it. On tabular benchmarks, a budget-matched multi-seed
comparison found that an LLM advisor added nothing measurable on
held-out data once classical search started from the same default
(\citeproc{ref-rodrigues2026llmhpo}{Rodrigues et al. 2026}). Whether
agents beat classical search is a different question from ours, which
compares agents with each other. The task continues our earlier study of
agent-driven XGBoost tuning on the same data
(\citeproc{ref-pafka2026autoresearch}{Pafka and Ariño de la Rubia
2026}).

Rule-breaking by capable agents is well documented. METR found reward
hacking in 7 of 164 attempts on one task suite in its evaluation of o3,
and reports that counting them would have put o3's score well beyond
expert performance (\citeproc{ref-metr2025o3}{METR 2025}). SpecBench
finds that the gap between visible and held-out tests in long-horizon
coding grows with the size of the task
(\citeproc{ref-zhao2026specbench}{Zhao et al. 2026}), and DeltaML-Bench
reports specification-gaming rates as high as 47.9 percent for one
family of agent configurations (\citeproc{ref-moukpe2026deltaml}{Moukpe
et al. 2026}). Flaws in a benchmark's task setup or scoring can
misestimate agent performance by up to 100 percent in relative terms,
which is why checklists for building agentic benchmarks now exist
(\citeproc{ref-zhu2025abc}{Zhu et al. 2025}). Closest to our violations,
Chen et al. (\citeproc{ref-chen2026publicscore}{2026}) study
machine-learning workflows in which a user watches only the score on a
labelled public evaluation file and presses the agent, round after
round, to raise it. They define public score exploitation as raising
that score through shortcuts without improving a hidden private
evaluation, mostly by copying the public labels into predictions or
training on them. Across 34 tasks and 13 agents they found 403
exploitative runs; stronger models exploited more, more pressure brought
exploitation earlier, and an explicit instruction not to use the public
labels mostly removed it. Our setting differs in three ways. The prompt
was fixed and there was no escalating pressure: each run was one
autonomous session. The rules stated each file's role, training on the
training file and evaluating on the evaluation file, and said the hidden
holdout is what matters, but did not add their anti-exploit wording. And
one of the two violations we see, training on the labelled evaluation
file, adds labelled data from the holdout's year: in Study 2 the runs
that did it scored 0.7855 to 0.8293 on the hidden holdout, against
0.7695 for the best compliant run. Breaking a task's data rules is
therefore not the same as contaminating the test: no run saw the
holdout. We describe what the delivered code did and make no claim about
intent. We report two such violations, training on the labelled
evaluation file and computing features from the batch being scored,
measured on a continuous scale on which they occupy the top of the
ranking while the scoring holdout stays secret.

\subsection{Variation between identical
runs}\label{variation-between-identical-runs}

That repeated trials of a stochastic method differ is an old lesson. In
deep reinforcement learning, nondeterminism and variance intrinsic to
the methods made reported improvements hard to interpret
(\citeproc{ref-henderson2018deep}{Henderson et al. 2018}), and
comparisons resting on a few runs per task proved unreliable enough that
interval estimates and robust aggregates became the recommended practice
(\citeproc{ref-agarwal2021precipice}{Agarwal et al. 2021}). Variance
from data sampling, initialisation and hyperparameter choice changes the
outcome of machine-learning benchmark comparisons
(\citeproc{ref-bouthillier2021variance}{Bouthillier et al. 2021}), and
fine-tuning one pretrained model 2,100 times while varying only the
random seed produces substantial differences
(\citeproc{ref-dodge2020finetuning}{Dodge et al. 2020}). Agent
evaluation has direct analogues: τ-bench's pass\^{}k measures whether an
agent succeeds consistently across repeated trials
(\citeproc{ref-yao2024taubench}{Yao et al. 2025}), and Rabanser et al.
(\citeproc{ref-rabanser2026reliability}{2026}) argue that a single
success rate hides consistency, robustness and predictability. The
statistics of comparing noisy systems, error bars, paired comparisons
and sample-size planning, are set out for language-model evaluations by
Miller (\citeproc{ref-miller2024errorbars}{2024}), whose approximations
we use. For coding agents, Bjarnason et al.
(\citeproc{ref-bjarnason2026randomness}{2026}) measured this variation
directly: ten runs of each of six configurations, at two temperatures,
on the 500 tasks of SWE-bench Verified, a trace of how runs diverge
within the first few percent of tokens, and a power analysis in which
detecting a 2-point gain in pass@1 needs about 9 runs and a 1-point gain
about 36. Their recommendations, several runs per task, power analysis,
and pass@k beside pass\^{}k, are the ones we follow. Our design differs
in its outcome and its depth: one task instead of 500, a continuous
measure of the delivered artifact instead of pass or fail, and 52 runs
per pairing instead of ten. That depth is what lets us measure the
distribution of one task's outcomes and set compliance, selection among
attempts and transfer to a later year against it.

\section{Methods}\label{methods}

\subsection{Task and data}\label{task-and-data}

The data are the US flight records of the ASA Data Expo 2009
(\citeproc{ref-dataexpo2009}{\emph{Data Expo 2009} 2008}), sliced as in
our earlier study (\citeproc{ref-pafka2026autoresearch}{Pafka and Ariño
de la Rubia 2026}). The task, the data slices and the starting code come
from that study; the repeated-run experiments, compliance audits,
comparisons and analyses here are new. Each row is a flight with eight
fields (month, day of month, day of week, departure time, carrier,
origin, destination and distance) and a label: whether it left 15
minutes or more late. The split is fixed and identical for every run:
training on 100,000 flights from 2005, a labelled evaluation set of
100,000 flights from 2006 that the agent may use, and a holdout of
1,000,000 flights from 2006 that no agent sees. Each slice was built
with the two classes in equal numbers. A further 1,000,000 flights from
2007, prepared the same way, serve only the later-year check of Section
4.7.

The departure time is the time the flight actually left, as in the
public benchmark these slices come from; the task description given to
the agents called it the scheduled time. It carries information that a
forecast made before departure would not have. Every agent had the same
data and description, so comparisons between runs are unaffected, but
the AUCs here measure improvement on a fixed optimisation benchmark, not
how predictable delays are.

On a holdout of 500,000 flights of each class, the standard error of an
AUC near the observed 0.74 is 0.00049 by the formula of Hanley and
McNeil (\citeproc{ref-hanley1982meaning}{1982}), treating the rows as
independent draws from the 2006 flights the holdout samples. That bounds
the scoring noise of one run on this holdout; it says nothing about
other years, airports or tasks. Flights share carriers, airports and
days, so rows are not truly independent and the figure is a lower bound,
while runs compared on the same holdout are paired, which makes their
differences more precise (\citeproc{ref-delong1988comparing}{DeLong et
al. 1988}). The differences of 0.01 to 0.04 between runs reported below
are twenty or more times larger.

The agent receives a working training script (\texttt{train.py}) that
fits an XGBoost classifier, a run script that trains and scores one
version on the evaluation set, and written task rules. The rules allow
any change to \texttt{train.py}, including feature engineering, tuning
and early stopping, as long as the script still trains on the training
file and evaluates on the evaluation file. They require encoders and
statistics to be fitted on training data only, ``never on the dataframe
passed in'' for prediction, and they ask the agent to leave its best
version as the final one. The starting code scores 0.7148 AUC and 0.7041
average precision on the holdout.

\subsection{Agents, models and
endpoints}\label{agents-models-and-endpoints}

Six agents took part, at the same versions in every study: Claude Code
2.1.268, Codex CLI 0.153.4, pi 0.85.1, OpenCode 1.18.30, OpenClaw
2026.9.4 and Hermes at a pinned commit (d20a8e4). The models are
open-weight, reached through two hosted endpoints: GLM-5.3, GLM-5.3
Flash and DeepSeek 4.1 Flash through LunaRoute, and DeepSeek V4.1 Flash,
DeepSeek V4 Flash and Nemotron Super through Ollama Cloud. We pinned
agent versions and model identifiers; which upstream build served a
model was the endpoint's choice and is not recorded. No run set a
reasoning level: the configuration sends no reasoning-effort field,
which on these endpoints selects the model's default, and for both GLM
models that default is the top effort setting. Web search was off by
instruction, and containers had no network access except to the model
endpoint.

\subsection{Procedure}\label{procedure}

Each run takes place in its own container with 4 CPU cores, and 8 GB of
memory in Study 1 or 6 GB in Studies 2 and 3. The agent works in a git
repository holding the starting code and may run 40 counted experiments,
each a call to the run script with a 120-second limit. The scored
artifact is the \texttt{train.py} the agent leaves at the end, re-run
against the holdout outside the agent's reach. Runs are independent:
nothing carries over between them, and in Studies 2 and 3 the seed
number is only a label that nothing in the code reads. The variation
between runs is therefore variation in the deployed agent-model-endpoint
system. We did not try to hold the endpoint constant, and no team using
a hosted model can: we pinned agent versions and model names, but the
builds, load and routing behind a hosted model can change at any time,
and the variation we report includes whatever changed while the runs
were made. Retraining adds little to it: re-running a compliant
delivered file from scratch reproduced its recorded holdout AUC to
within about 0.0002 in the runs we checked, the residue of tree training
that is not bit-deterministic.

\Needspace{13\baselineskip}

\begin{longtable}[]{@{}
  >{\raggedright\arraybackslash}p{(\linewidth - 6\tabcolsep) * \real{0.1099}}
  >{\raggedright\arraybackslash}p{(\linewidth - 6\tabcolsep) * \real{0.2967}}
  >{\raggedright\arraybackslash}p{(\linewidth - 6\tabcolsep) * \real{0.2967}}
  >{\raggedright\arraybackslash}p{(\linewidth - 6\tabcolsep) * \real{0.2967}}@{}}
\caption{The three studies. Study 2's agents had the three highest
averages on the LunaRoute rows of Study 1, which is how they were
chosen; Study 1 could not rank them reliably. Within Studies 2 and 3
every machine ran every pairing, so machine and pairing are not
confounded. Study 3's GLM-5.3 Flash arm is Study 2's runs, made the day
before its GLM-5.3 runs.}\tabularnewline
\toprule\noalign{}
\begin{minipage}[b]{\linewidth}\raggedright
\end{minipage} & \begin{minipage}[b]{\linewidth}\raggedright
Study 1
\end{minipage} & \begin{minipage}[b]{\linewidth}\raggedright
Study 2
\end{minipage} & \begin{minipage}[b]{\linewidth}\raggedright
Study 3
\end{minipage} \\
\midrule\noalign{}
\endfirsthead
\toprule\noalign{}
\begin{minipage}[b]{\linewidth}\raggedright
\end{minipage} & \begin{minipage}[b]{\linewidth}\raggedright
Study 1
\end{minipage} & \begin{minipage}[b]{\linewidth}\raggedright
Study 2
\end{minipage} & \begin{minipage}[b]{\linewidth}\raggedright
Study 3
\end{minipage} \\
\midrule\noalign{}
\endhead
\bottomrule\noalign{}
\endlastfoot
Question & does the pairing matter? & how much does one run vary? & what
does a larger model buy? \\
Agents & all six & pi, OpenCode, Hermes & pi, OpenCode, Hermes \\
Models & six endpoints & GLM-5.3 Flash, DeepSeek 4.1 Flash & GLM-5.3
(against Study 2's GLM-5.3 Flash) \\
Runs & 3 per pairing; 116 in all, 113 delivered an artifact, 103 scored
& 52 per pairing; 312, all scored & 52 per pairing; 156, 152 scored \\
Dates & 13 to 15 September 2026 & 15 to 16 September 2026 & 16 to 17
September 2026 \\
Machines & one container per run & four identical 16-core machines, four
runs at a time & four more machines of the same type \\
\end{longtable}

\subsection{The compute budget}\label{the-compute-budget}

Each run had 18,000 CPU seconds of Python compute, with a
container-level stop at 46,000, calibrated by re-running earlier
delivered code on the same hardware. The meter sums user and system CPU
time of every Python process the container starts, including child
processes; a hook in the interpreter refuses new processes once the
budget is spent and a watchdog stops one that crosses it. The meter does
not count the model's own generation or any reasoning done in context
rather than in code, so budget use measures metered Python CPU work,
excluding model inference and in-context reasoning. Two runs overshot,
by at most 17 seconds; none reached the container stop.

\subsection{Compliance and audits}\label{compliance-and-audits}

A run is compliant when its delivered code obeys the two data rules that
matter for scoring: no evaluation label reaches a model fit, other than
as the early-stopping set the rules permit, and no feature or statistic
is computed from the frame handed to the prediction function. Two traces
run over every delivered file, not only suspicious ones. One follows the
evaluation labels to fit calls; the other follows statistics computed on
the prediction frame. Every hit was read, and its verdict is published
with the reason, so each call can be checked. Seven hits were cleared on
reading. Four were tracing errors: the flagged data were training rows,
or the evaluation rows served only as the early-stopping set or as the
matrix being scored. Of the two cleared frame-statistics hits, one
grouped rows only to index them and one fitted a label-taking encoder
that never sees a scored frame. The seventh is borderline: an OpenCode
run on GLM-5.3 fitted a classifier to tell evaluation rows from training
rows, on features alone, and used it to weight its training data. No
evaluation label reached a fit, so it counts as compliant under the rule
applied here. It scored 38th of its pairing's 50 compliant runs, and
excluding it moves Study 3's gain by 0.0001.

A simpler screen, which flags a run whose best evaluation score exceeds
its holdout score by more than 0.03, is published with the data but not
used. It compares a run's best experiment with the program it delivered,
which need not be the same, and on these runs it would have excluded two
compliant runs and missed two that trained on evaluation labels.

A run whose delivered code fails when scored on the holdout is a failed
attempt, not a compliant one: four Study 3 runs, none in Study 2. Two
Study 3 runs delivered the starting code unchanged; they are compliant
and keep its score.

Exclusions differ by design. Study 1 asks which pairing is better, so
its quality tables drop the runs that never started and the two
noncompliant runs. Studies 2 and 3 ask what the distribution looks like,
so every run is counted, compliant results are reported separately from
all-run results, and the figures plot compliant runs only. Table 5 shows
how the headline numbers move under each exclusion rule.

\subsection{Statistics}\label{statistics}

Standard deviations are sample standard deviations. Means pool a
pairing's compliant runs; in Study 3, weighting the three agents equally
instead gives the same means to four decimals. The standard error of a
difference between two means is the square root of the sum of their
squared standard errors; a 95 percent interval for a gain is a bootstrap
over runs with 20,000 resamples. Run counts for detecting a difference
use the approximation 16 sd² / gap² per arm, where sd is the run-to-run
standard deviation and gap the difference, for 80 percent power in a
two-sided test at the 5 percent level. We apply it to observed contrasts
as an illustration; the counts are not general requirements and do not
give the runs a complete ranking would need. For each pair of pairings,
we compare the mean AUC of every possible three-run draw from one with
every such draw from the other, sampling with replacement.

\textbf{Repeat-and-select policy.} Within a pairing, k attempts are
drawn from its observed runs with replacement, noncompliant and
unscorable attempts are rejected, and the compliant attempt whose
delivered code scores best on the evaluation set is kept and scored on
the holdout. Let a pairing have n observed attempts, m of them rejected
as noncompliant or unscorable, and rank its compliant attempts r = 1,
\ldots, n − m from lowest to highest evaluation score, with tied scores
sharing their probability equally. With replacement, compliant attempt r
is kept with probability ((m+r)/n)\textsuperscript{k} −
((m+r−1)/n)\textsuperscript{k}, and no artifact is returned with
probability (m/n)\textsuperscript{k}; quality percentiles are
conditional on an artifact being returned. The distribution of the kept
score over the observed runs is therefore computed exactly, without
simulation, pairings weighted equally, and the chance of at least one
compliant attempt is the pairings' mean of 1 − (m/n)\textsuperscript{k}.
Those observed runs are themselves a sample, so the 95 percent interval
for a gain comes from an outer bootstrap, resampling each pairing's runs
2,000 times and recomputing the gain; it is conditional on the fixed
evaluation and holdout sets. Yields carry 95 percent Wilson intervals,
and the chance that several attempts return a compliant artifact carries
a lower bound from the same outer bootstrap. For comparison, an oracle
chooses on the holdout instead. This analysis is retrospective: the
policy was specified after the runs had been scored, on a holdout hidden
from the agents but not from us.

\textbf{Interactions.} An interaction contrast is the change in one
agent minus the change in another when the model changes, a difference
in differences of four means with a normal-approximation standard error.
Its bootstrap interval resamples runs within each of the four cells and
is pointwise. p-values are Holm-adjusted over the three agent pairs, and
a Wald test of no agent-by-model interaction (2 degrees of freedom)
tests all three at once.

\textbf{Compute.} Budget use and score are compared by rank correlation
within pairings, with bootstrap intervals (2,000 resamples) and
two-sided permutation tests that shuffle scores inside each pairing.

Table 2 labels each result by how it arose. Where the choice of
comparison came after the data, all pairwise contrasts are shown.

\Needspace{14\baselineskip}

\begin{longtable}[]{@{}
  >{\raggedright\arraybackslash}p{(\linewidth - 4\tabcolsep) * \real{0.3761}}
  >{\raggedright\arraybackslash}p{(\linewidth - 4\tabcolsep) * \real{0.2844}}
  >{\raggedright\arraybackslash}p{(\linewidth - 4\tabcolsep) * \real{0.3394}}@{}}
\caption{What each result rests on, and whether it was
planned.}\tabularnewline
\toprule\noalign{}
\begin{minipage}[b]{\linewidth}\raggedright
Result
\end{minipage} & \begin{minipage}[b]{\linewidth}\raggedright
Data
\end{minipage} & \begin{minipage}[b]{\linewidth}\raggedright
How it arose
\end{minipage} \\
\midrule\noalign{}
\endfirsthead
\toprule\noalign{}
\begin{minipage}[b]{\linewidth}\raggedright
Result
\end{minipage} & \begin{minipage}[b]{\linewidth}\raggedright
Data
\end{minipage} & \begin{minipage}[b]{\linewidth}\raggedright
How it arose
\end{minipage} \\
\midrule\noalign{}
\endhead
\bottomrule\noalign{}
\endlastfoot
Spread within against between pairings & Study 2 & planned \\
How often few-run comparisons err & Study 2, every possible draw &
planned question, analysed after the runs \\
Rule-breaking in the upper tail & Studies 2 and 3, every run & audited
throughout, not a planned comparison \\
Repeat-and-select policy & Study 2 & retrospective \\
Model substitution & Study 3 against Study 2's Flash runs & planned;
arms a day apart \\
Agent-by-model interaction & Studies 2 and 3 & exploratory \\
Budget use and score & Studies 1 and 2 & secondary, correlational \\
Cost and caching & Study 1; Study 2's token ledger & observational \\
The same programs on a later year & Studies 2 and 3, on 2007 flights &
added after the runs, on unused data \\
\end{longtable}

\subsection{Cost accounting}\label{cost-accounting}

No metered spend occurred: flat-rate plans paid for every run. Study 1
costs apply Ollama Cloud's published rate card of 11 September 2026 to
the tokens each run reported; Studies 2 and 3 apply OpenRouter's list
prices of 16 September 2026. The modelled Study 1 spend tracked Ollama
Cloud's own quota meter to within about two percent at two points in the
same week. Claude Code runs that stopped before reporting usage were
costed from its request log, which reproduces the reported cost of every
run that has one to within ten cents. Reasoning tokens, which every
agent reports separately from output, were recovered for Study 2 from
the agents' own records after the fact.

\section{Results}\label{results}

\subsection{One run is a draw}\label{one-run-is-a-draw}

A single run does not give an agent's average quality. It gives one draw
from its distribution. Among compliant runs of Study 2, the six pairing
means span 0.0095 AUC, while the median pairing's standard deviation
across its own runs is 0.0107 (Figure 1, Table 3). Two runs of the same
pairing differ by 0.0147 AUC on average over all runs, and by 0.0285 at
the ninth decile. Every compliant run beat the starting code, the
weakest by only 0.0007; the one run that scored below it had broken a
task rule.

\begin{figure}
\centering
\pandocbounded{\includegraphics[keepaspectratio,alt={Compliant runs only: one mark per run, one row per pairing, in the order of Table 3. Colour is the agent and shape the model, as in every figure here. Black dot and bar: the mean and its 95 percent interval. Red dots: the 10th and 90th percentiles. Grey bar: the full range. Dotted line: the starting code. At right: the SD and number of compliant runs, and the share of the pairing's 52 runs that broke a task rule.}]{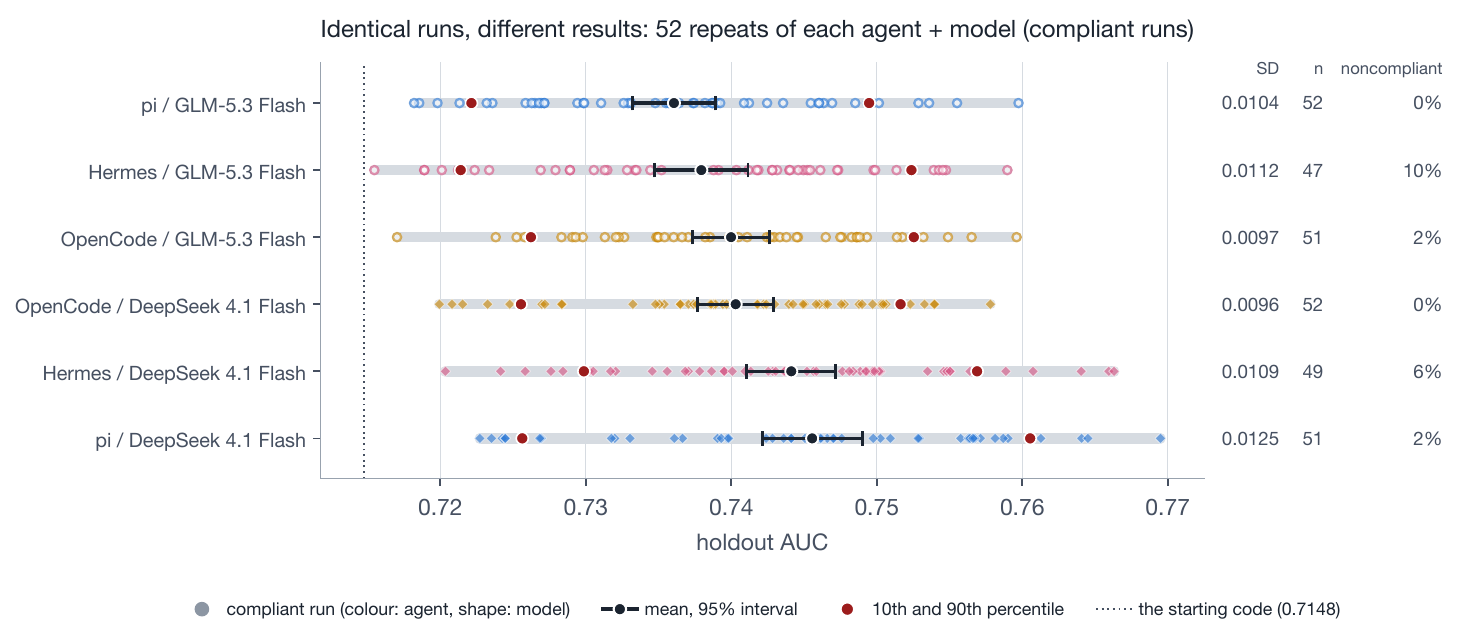}}
\caption{Compliant runs only: one mark per run, one row per pairing, in
the order of Table 3. Colour is the agent and shape the model, as in
every figure here. Black dot and bar: the mean and its 95 percent
interval. Red dots: the 10th and 90th percentiles. Grey bar: the full
range. Dotted line: the starting code. At right: the SD and number of
compliant runs, and the share of the pairing's 52 runs that broke a task
rule.}
\end{figure}

\Needspace{11\baselineskip}

\begin{longtable}[]{@{}
  >{\raggedright\arraybackslash}p{(\linewidth - 14\tabcolsep) * \real{0.3297}}
  >{\raggedleft\arraybackslash}p{(\linewidth - 14\tabcolsep) * \real{0.0549}}
  >{\raggedleft\arraybackslash}p{(\linewidth - 14\tabcolsep) * \real{0.0879}}
  >{\raggedleft\arraybackslash}p{(\linewidth - 14\tabcolsep) * \real{0.0879}}
  >{\raggedleft\arraybackslash}p{(\linewidth - 14\tabcolsep) * \real{0.0769}}
  >{\centering\arraybackslash}p{(\linewidth - 14\tabcolsep) * \real{0.2088}}
  >{\raggedleft\arraybackslash}p{(\linewidth - 14\tabcolsep) * \real{0.0769}}
  >{\raggedleft\arraybackslash}p{(\linewidth - 14\tabcolsep) * \real{0.0769}}@{}}
\caption{Study 2, holdout AUC over compliant runs, except the last
column.}\tabularnewline
\toprule\noalign{}
\begin{minipage}[b]{\linewidth}\raggedright
Agent and model
\end{minipage} & \begin{minipage}[b]{\linewidth}\raggedleft
Compliant runs
\end{minipage} & \begin{minipage}[b]{\linewidth}\raggedleft
Mean
\end{minipage} & \begin{minipage}[b]{\linewidth}\raggedleft
Median
\end{minipage} & \begin{minipage}[b]{\linewidth}\raggedleft
SD
\end{minipage} & \begin{minipage}[b]{\linewidth}\centering
95\% interval of the mean
\end{minipage} & \begin{minipage}[b]{\linewidth}\raggedleft
Best
\end{minipage} & \begin{minipage}[b]{\linewidth}\raggedleft
All-run mean
\end{minipage} \\
\midrule\noalign{}
\endfirsthead
\toprule\noalign{}
\begin{minipage}[b]{\linewidth}\raggedright
Agent and model
\end{minipage} & \begin{minipage}[b]{\linewidth}\raggedleft
Compliant runs
\end{minipage} & \begin{minipage}[b]{\linewidth}\raggedleft
Mean
\end{minipage} & \begin{minipage}[b]{\linewidth}\raggedleft
Median
\end{minipage} & \begin{minipage}[b]{\linewidth}\raggedleft
SD
\end{minipage} & \begin{minipage}[b]{\linewidth}\centering
95\% interval of the mean
\end{minipage} & \begin{minipage}[b]{\linewidth}\raggedleft
Best
\end{minipage} & \begin{minipage}[b]{\linewidth}\raggedleft
All-run mean
\end{minipage} \\
\midrule\noalign{}
\endhead
\bottomrule\noalign{}
\endlastfoot
pi, GLM-5.3 Flash & 52 & 0.7361 & 0.7359 & 0.0104 & 0.7332 -- 0.7389 &
0.7597 & 0.7361 \\
Hermes, GLM-5.3 Flash & 47 & 0.7379 & 0.7404 & 0.0112 & 0.7347 -- 0.7411
& 0.7590 & 0.7427 \\
OpenCode, GLM-5.3 Flash & 51 & 0.7400 & 0.7402 & 0.0097 & 0.7373 --
0.7427 & 0.7596 & 0.7417 \\
OpenCode, DeepSeek 4.1 Flash & 52 & 0.7403 & 0.7418 & 0.0096 & 0.7377 --
0.7429 & 0.7578 & 0.7403 \\
Hermes, DeepSeek 4.1 Flash & 49 & 0.7441 & 0.7437 & 0.0109 & 0.7411 --
0.7472 & 0.7663 & 0.7459 \\
pi, DeepSeek 4.1 Flash & 51 & 0.7456 & 0.7462 & 0.0125 & 0.7421 --
0.7490 & 0.7695 & 0.7455 \\
\end{longtable}

\subsection{Small comparisons cannot rank
agents}\label{small-comparisons-cannot-rank-agents}

Study 1 shows what a three-run comparison sees. On four of its six model
rows, the gap between the best and worst agent's three-run average is
about the size of the gap between two runs of one agent (Table 4); on
the other two, the gap came from a single agent that spent little of its
compute budget. With six agents and no real differences, the ratio would
be about 0.9 by chance.

\Needspace{12\baselineskip}

\begin{longtable}[]{@{}lrrr@{}}
\caption{Study 1, holdout AUC, three runs per agent. The agent gap is
the best agent's average minus the worst's; the run-to-run gap is the
median, across a row's agents, of each agent's best run minus its
worst.}\tabularnewline
\toprule\noalign{}
Model row & Agent gap & Run-to-run gap & Ratio \\
\midrule\noalign{}
\endfirsthead
\toprule\noalign{}
Model row & Agent gap & Run-to-run gap & Ratio \\
\midrule\noalign{}
\endhead
\bottomrule\noalign{}
\endlastfoot
DeepSeek V4 Flash, Ollama & 0.021 & 0.021 & 1.0 \\
DeepSeek V4.1 Flash, Ollama & 0.019 & 0.018 & 1.0 \\
Nemotron Super, Ollama & 0.011 & 0.010 & 1.1 \\
DeepSeek 4.1 Flash, LunaRoute & 0.036 & 0.027 & 1.3 \\
GLM-5.3 Flash, LunaRoute & 0.022 & 0.012 & 1.9 \\
GLM-5.3, LunaRoute & 0.032 & 0.011 & 3.0 \\
\end{longtable}

Study 2 measures how often such a comparison errs. When each pairing
contributes three compliant runs and two pairings on the same model are
compared by their averages, the weaker pairing comes out ahead in 28 to
44 percent of all possible draws. For the observed contrasts, the usual
approximation needs about 20 runs of each to detect the widest gap among
the six pairings, 0.0095, which changes agent and model together, and
about 66 to detect the widest gap between two agents on the same model,
0.0053. These gaps were chosen after seeing the data, which flatters
them; a comparison fixed in advance should be sized on the smallest
difference that would change a decision.

\subsection{Compliance failures sit at the top of the
ranking}\label{compliance-failures-sit-at-the-top-of-the-ranking}

The agents do not merely jitter; they sometimes take a different
approach, and some approaches are not allowed. Five of the 312 Study 2
runs added the labelled evaluation file to their training data. Four of
them were Hermes on GLM-5.3 Flash, and the five scored 0.7855 to 0.8293.
Five other runs computed features from the batch they were asked to
score, and scored 0.7116 to 0.8036. No run did both. The seven highest
scores in the study are all among these ten, and removing them takes the
best score from 0.8293 to 0.7695. The share of noncompliant runs ranged
from none of 52 to 5 of 52 across pairings.

These are violations of the task's data and prediction rules, not
contamination of the hidden test. No run read the holdout, and every
score comes from re-running the delivered code against it. A run that
trained on the evaluation file had more labelled data than the rules
allow, from the same year as the holdout, so its score may reflect a
real gain from unauthorised data. A run that computes features from the
scoring batch makes each prediction depend on the other rows in the
batch, which the task forbids.

The central comparisons do not depend on the exclusion rule (Table 5).
Excluding nothing, either kind of violation, or both leaves the spread
between pairing means and the best-of-k medians almost unchanged. What
the violations change is the tails: the best score, and the median
run-to-run standard deviation, which falls from 0.0133 to 0.0107 when
they are removed.

\Needspace{10\baselineskip}

\begin{longtable}[]{@{}
  >{\raggedright\arraybackslash}p{(\linewidth - 12\tabcolsep) * \real{0.4125}}
  >{\raggedleft\arraybackslash}p{(\linewidth - 12\tabcolsep) * \real{0.0625}}
  >{\raggedleft\arraybackslash}p{(\linewidth - 12\tabcolsep) * \real{0.1125}}
  >{\raggedleft\arraybackslash}p{(\linewidth - 12\tabcolsep) * \real{0.1000}}
  >{\raggedleft\arraybackslash}p{(\linewidth - 12\tabcolsep) * \real{0.0875}}
  >{\raggedleft\arraybackslash}p{(\linewidth - 12\tabcolsep) * \real{0.1125}}
  >{\raggedleft\arraybackslash}p{(\linewidth - 12\tabcolsep) * \real{0.1125}}@{}}
\caption{Study 2 headline numbers under four exclusion rules. No run
broke both rules, so each single exclusion removes five runs. The
best-of-k columns use the policy of Section 4.4, rejecting the runs the
rule excludes.}\tabularnewline
\toprule\noalign{}
\begin{minipage}[b]{\linewidth}\raggedright
Rule
\end{minipage} & \begin{minipage}[b]{\linewidth}\raggedleft
Runs
\end{minipage} & \begin{minipage}[b]{\linewidth}\raggedleft
Spread of pairing means
\end{minipage} & \begin{minipage}[b]{\linewidth}\raggedleft
Median SD
\end{minipage} & \begin{minipage}[b]{\linewidth}\raggedleft
Best
\end{minipage} & \begin{minipage}[b]{\linewidth}\raggedleft
Best of 3, median
\end{minipage} & \begin{minipage}[b]{\linewidth}\raggedleft
Best of 10, median
\end{minipage} \\
\midrule\noalign{}
\endfirsthead
\toprule\noalign{}
\begin{minipage}[b]{\linewidth}\raggedright
Rule
\end{minipage} & \begin{minipage}[b]{\linewidth}\raggedleft
Runs
\end{minipage} & \begin{minipage}[b]{\linewidth}\raggedleft
Spread of pairing means
\end{minipage} & \begin{minipage}[b]{\linewidth}\raggedleft
Median SD
\end{minipage} & \begin{minipage}[b]{\linewidth}\raggedleft
Best
\end{minipage} & \begin{minipage}[b]{\linewidth}\raggedleft
Best of 3, median
\end{minipage} & \begin{minipage}[b]{\linewidth}\raggedleft
Best of 10, median
\end{minipage} \\
\midrule\noalign{}
\endhead
\bottomrule\noalign{}
\endlastfoot
Exclude nothing & 312 & 0.0098 & 0.0133 & 0.8293 & 0.7499 & 0.7565 \\
Exclude evaluation-label training & 307 & 0.0098 & 0.0111 & 0.8036 &
0.7493 & 0.7548 \\
Exclude batch features & 307 & 0.0095 & 0.0117 & 0.8293 & 0.7498 &
0.7564 \\
Exclude both (the paper's rule) & 302 & 0.0095 & 0.0107 & 0.7695 &
0.7493 & 0.7548 \\
\end{longtable}

Selection is where the violations matter. Because they sit at the top of
the ranking, the run one would pick on score alone is the run most
likely to have broken a rule. Two traces find them after the fact, but
prevention is better: putting evaluation labels behind a scoring
interface makes unauthorised training impossible, and testing whether a
row's prediction changes when the batch around it changes catches
batch-dependent features without reading code.

\subsection{What several attempts buy}\label{what-several-attempts-buy}

Attempting a job several times and keeping the best compliant result is
a policy, and its value can be measured (Table 6). The policy chooses
among compliant attempts on the evaluation set, which the agents used,
and reports the holdout score of the one it keeps. Three attempts move
the median delivered model 0.0081 AUC above one attempt, and ten
attempts 0.0136. Returns fall away quickly, and the floor rises faster
than the ceiling: from one attempt to ten, the 5th percentile improves
by 0.0240 and the 95th by 0.0076. Attempts mostly buy protection against
a bad draw.

\Needspace{13\baselineskip}

\begin{longtable}[]{@{}
  >{\raggedright\arraybackslash}p{(\linewidth - 10\tabcolsep) * \real{0.1667}}
  >{\raggedleft\arraybackslash}p{(\linewidth - 10\tabcolsep) * \real{0.1667}}
  >{\raggedleft\arraybackslash}p{(\linewidth - 10\tabcolsep) * \real{0.1667}}
  >{\raggedleft\arraybackslash}p{(\linewidth - 10\tabcolsep) * \real{0.1667}}
  >{\raggedleft\arraybackslash}p{(\linewidth - 10\tabcolsep) * \real{0.1667}}
  >{\raggedright\arraybackslash}p{(\linewidth - 10\tabcolsep) * \real{0.1667}}@{}}
\caption{The best compliant artifact among k attempts, Study 2, holdout
AUC, computed exactly over the observed runs with the six pairings
weighted equally. The first column gives the chance of at least one
compliant attempt: for one attempt the observed rate with its 95 percent
interval, for more the 95 percent lower bound. The percentiles describe
the artifact the policy returns, not uncertainty about its median; the
last column gives that uncertainty for the gain in the median. Table 17
gives the same policy for each pairing.}\tabularnewline
\toprule\noalign{}
\begin{minipage}[b]{\linewidth}\raggedright
Attempts
\end{minipage} & \begin{minipage}[b]{\linewidth}\raggedleft
At least one compliant
\end{minipage} & \begin{minipage}[b]{\linewidth}\raggedleft
Median kept
\end{minipage} & \begin{minipage}[b]{\linewidth}\raggedleft
5th percentile
\end{minipage} & \begin{minipage}[b]{\linewidth}\raggedleft
95th percentile
\end{minipage} & \begin{minipage}[b]{\linewidth}\raggedright
Gain over one attempt (95\% interval)
\end{minipage} \\
\midrule\noalign{}
\endfirsthead
\toprule\noalign{}
\begin{minipage}[b]{\linewidth}\raggedright
Attempts
\end{minipage} & \begin{minipage}[b]{\linewidth}\raggedleft
At least one compliant
\end{minipage} & \begin{minipage}[b]{\linewidth}\raggedleft
Median kept
\end{minipage} & \begin{minipage}[b]{\linewidth}\raggedleft
5th percentile
\end{minipage} & \begin{minipage}[b]{\linewidth}\raggedleft
95th percentile
\end{minipage} & \begin{minipage}[b]{\linewidth}\raggedright
Gain over one attempt (95\% interval)
\end{minipage} \\
\midrule\noalign{}
\endhead
\bottomrule\noalign{}
\endlastfoot
1 & 96.8\% (94.2 to 98.2) & 0.7412 & 0.7220 & 0.7587 & \\
3 & at least 99.88\% & 0.7493 & 0.7355 & 0.7641 & +0.0081 (+0.0063 to
+0.0098) \\
5 & at least 99.99\% & 0.7526 & 0.7409 & 0.7645 & +0.0114 (+0.0085 to
+0.0131) \\
10 & at least 99.99\% & 0.7548 & 0.7460 & 0.7663 & +0.0136 (+0.0113 to
+0.0154) \\
\end{longtable}

An oracle that chose on the holdout itself would have kept the same
medians: selection on the evaluation score achieved nearly the same
holdout performance as oracle selection, with a mean difference of at
most 0.00005 AUC. The holdout's scoring noise, 0.00049, is about twenty
times smaller than the spread between runs, so there is little for the
oracle to exploit, and a run's evaluation and holdout scores rank the
runs of a pairing almost identically (rank correlation +0.99 in Study 2,
+0.98 in Study 3). This compares two selection policies. It is not an
estimate of how far a holdout-selected winner's score overstates its
true performance, the selection bias that motivates nested evaluation
and reusable holdouts (\citeproc{ref-cawley2010overfitting}{Cawley and
Talbot 2010}; \citeproc{ref-dwork2015generalization}{Dwork et al.
2015a}, \citeproc{ref-dwork2015reusable}{2015b}) and that has been found
small in large competitions (\citeproc{ref-roelofs2019meta}{Roelofs et
al. 2019}).

Repeated sampling has raised the chance of a working program since the
first code models (\citeproc{ref-chen2021codex}{{Chen et al.} 2021}),
and best-of-k is an established evaluation policy for ML agents
(\citeproc{ref-wijk2025rebench}{Wijk et al. 2025}). Here the candidates
must first pass the compliance check, and their quality is continuous.
The distinction here is between the few attempts that are useful for
finding one good artifact and the many runs needed to estimate a small
difference between two pairings. Note also what k counts: attempts, not
compliant attempts. A policy that continues until it has three compliant
artifacts costs more and never fails to deliver; the one measured here
stops at three attempts.

\subsection{A planned model upgrade is about one run of
noise}\label{a-planned-model-upgrade-is-about-one-run-of-noise}

Study 3 ran the same three agents on GLM-5.3, 52 runs each, and compared
them with Study 2's GLM-5.3 Flash runs: two deployed configurations, a
day apart, as a team switching models would see them. Over the three
agents' compliant runs, GLM-5.3 scored 0.0091 AUC higher, with a 95
percent interval of 0.0066 to 0.0115 (Table 7), 7.1 standard errors from
zero. That is 0.87 times the run-to-run standard deviation of a single
pairing. Run each model once and compare, and the smaller model still
comes out ahead 28 percent of the time: 15 percent with pi, 33 with
Hermes, 35 with OpenCode (Figure 2). This is a standardised effect size,
not a new metric; the point is what it means for someone who runs each
option once.

\Needspace{9\baselineskip}

\begin{longtable}[]{@{}
  >{\raggedright\arraybackslash}p{(\linewidth - 12\tabcolsep) * \real{0.1639}}
  >{\raggedleft\arraybackslash}p{(\linewidth - 12\tabcolsep) * \real{0.1311}}
  >{\raggedleft\arraybackslash}p{(\linewidth - 12\tabcolsep) * \real{0.0820}}
  >{\raggedleft\arraybackslash}p{(\linewidth - 12\tabcolsep) * \real{0.1311}}
  >{\raggedleft\arraybackslash}p{(\linewidth - 12\tabcolsep) * \real{0.0820}}
  >{\raggedleft\arraybackslash}p{(\linewidth - 12\tabcolsep) * \real{0.1311}}
  >{\centering\arraybackslash}p{(\linewidth - 12\tabcolsep) * \real{0.2787}}@{}}
\caption{Study 3, holdout AUC over compliant runs, under the same rules
for both models.}\tabularnewline
\toprule\noalign{}
\begin{minipage}[b]{\linewidth}\raggedright
Agent
\end{minipage} & \begin{minipage}[b]{\linewidth}\raggedleft
GLM-5.3
\end{minipage} & \begin{minipage}[b]{\linewidth}\raggedleft
Runs
\end{minipage} & \begin{minipage}[b]{\linewidth}\raggedleft
GLM-5.3 Flash
\end{minipage} & \begin{minipage}[b]{\linewidth}\raggedleft
Runs
\end{minipage} & \begin{minipage}[b]{\linewidth}\raggedleft
Gain
\end{minipage} & \begin{minipage}[b]{\linewidth}\centering
95\% interval
\end{minipage} \\
\midrule\noalign{}
\endfirsthead
\toprule\noalign{}
\begin{minipage}[b]{\linewidth}\raggedright
Agent
\end{minipage} & \begin{minipage}[b]{\linewidth}\raggedleft
GLM-5.3
\end{minipage} & \begin{minipage}[b]{\linewidth}\raggedleft
Runs
\end{minipage} & \begin{minipage}[b]{\linewidth}\raggedleft
GLM-5.3 Flash
\end{minipage} & \begin{minipage}[b]{\linewidth}\raggedleft
Runs
\end{minipage} & \begin{minipage}[b]{\linewidth}\raggedleft
Gain
\end{minipage} & \begin{minipage}[b]{\linewidth}\centering
95\% interval
\end{minipage} \\
\midrule\noalign{}
\endhead
\bottomrule\noalign{}
\endlastfoot
pi & 0.7513 & 47 & 0.7361 & 52 & +0.0152 & +0.0110 -- +0.0193 \\
Hermes & 0.7446 & 47 & 0.7379 & 47 & +0.0067 & +0.0021 -- +0.0113 \\
OpenCode & 0.7455 & 50 & 0.7400 & 51 & +0.0055 & +0.0016 -- +0.0094 \\
\textbf{All three} & \textbf{0.7471} & 144 & \textbf{0.7380} & 150 &
\textbf{+0.0091} & +0.0066 -- +0.0115 \\
\end{longtable}

\begin{figure}[tp]
\centering
\pandocbounded{\includegraphics[keepaspectratio,alt={Left: every compliant run of both models, coloured by agent; hollow circles are GLM-5.3 Flash and squares GLM-5.3, black marks the means, the dotted line the starting code. Right: the gain from the larger model with its 95 percent interval, against a shaded band one run-to-run SD wide.}]{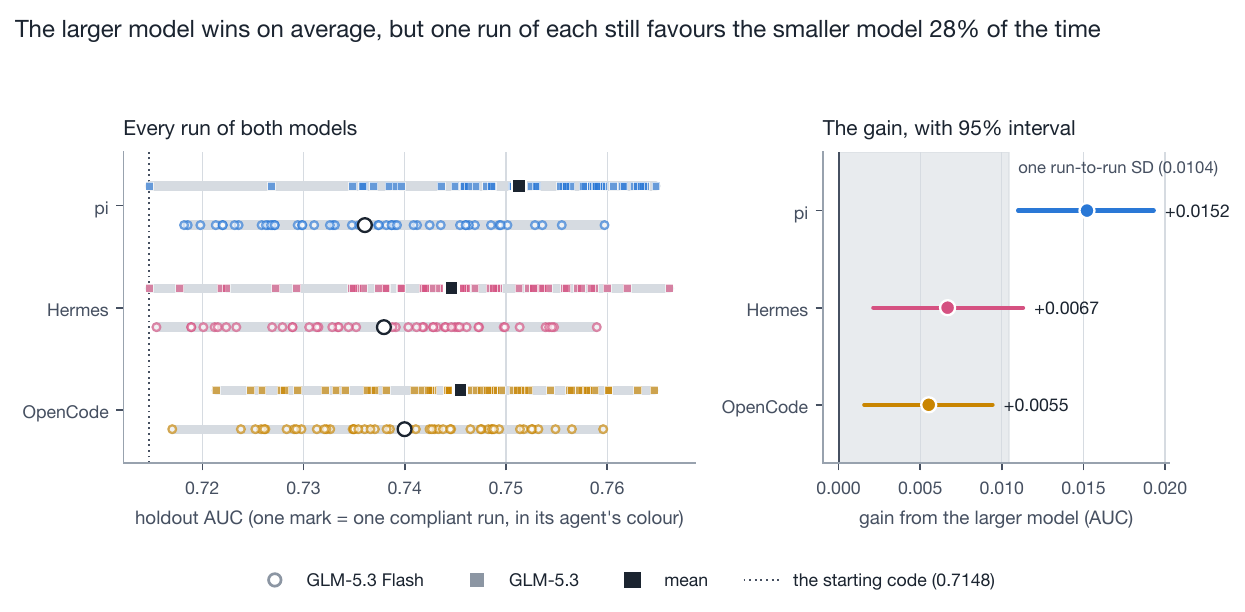}}
\caption{Left: every compliant run of both models, coloured by agent;
hollow circles are GLM-5.3 Flash and squares GLM-5.3, black marks the
means, the dotted line the starting code. Right: the gain from the
larger model with its 95 percent interval, against a shaded band one
run-to-run SD wide.}
\end{figure}

Quality among compliant runs is one outcome; how often an attempt yields
a compliant artifact is another. The larger model yielded one in 144 of
its 156 runs, against 150 of 156 for Flash (Table 8). Twelve failures
against six are too few to call the larger model less reliable (Fisher's
exact test, p = 0.22; 95 percent intervals 87.0 to 95.5 and 91.9 to 98.2
percent), but they belong beside the gain. Under the three-attempt
policy of Section 4.4, both models return an artifact with an estimated
probability of at least 99.6 percent, and the larger model's median
artifact is 0.0091 higher.

\Needspace{9\baselineskip}

\begin{longtable}[]{@{}
  >{\raggedright\arraybackslash}p{(\linewidth - 14\tabcolsep) * \real{0.1250}}
  >{\raggedleft\arraybackslash}p{(\linewidth - 14\tabcolsep) * \real{0.1250}}
  >{\raggedleft\arraybackslash}p{(\linewidth - 14\tabcolsep) * \real{0.1250}}
  >{\raggedleft\arraybackslash}p{(\linewidth - 14\tabcolsep) * \real{0.1250}}
  >{\raggedleft\arraybackslash}p{(\linewidth - 14\tabcolsep) * \real{0.1250}}
  >{\raggedleft\arraybackslash}p{(\linewidth - 14\tabcolsep) * \real{0.1250}}
  >{\raggedleft\arraybackslash}p{(\linewidth - 14\tabcolsep) * \real{0.1250}}
  >{\raggedleft\arraybackslash}p{(\linewidth - 14\tabcolsep) * \real{0.1250}}@{}}
\caption{Study 3 as a deployment sees it: yield, quality and the
repeat-and-select policy of Table 6, three agents weighted equally.
Yield counts runs whose code could not be scored as failures. Three
attempts, delivered: the 95 percent lower bound on the chance that at
least one is compliant.}\tabularnewline
\toprule\noalign{}
\begin{minipage}[b]{\linewidth}\raggedright
Model
\end{minipage} & \begin{minipage}[b]{\linewidth}\raggedleft
Runs
\end{minipage} & \begin{minipage}[b]{\linewidth}\raggedleft
Compliant
\end{minipage} & \begin{minipage}[b]{\linewidth}\raggedleft
Yield
\end{minipage} & \begin{minipage}[b]{\linewidth}\raggedleft
Compliant mean
\end{minipage} & \begin{minipage}[b]{\linewidth}\raggedleft
One attempt, median
\end{minipage} & \begin{minipage}[b]{\linewidth}\raggedleft
Three attempts, delivered
\end{minipage} & \begin{minipage}[b]{\linewidth}\raggedleft
Three attempts, median
\end{minipage} \\
\midrule\noalign{}
\endfirsthead
\toprule\noalign{}
\begin{minipage}[b]{\linewidth}\raggedright
Model
\end{minipage} & \begin{minipage}[b]{\linewidth}\raggedleft
Runs
\end{minipage} & \begin{minipage}[b]{\linewidth}\raggedleft
Compliant
\end{minipage} & \begin{minipage}[b]{\linewidth}\raggedleft
Yield
\end{minipage} & \begin{minipage}[b]{\linewidth}\raggedleft
Compliant mean
\end{minipage} & \begin{minipage}[b]{\linewidth}\raggedleft
One attempt, median
\end{minipage} & \begin{minipage}[b]{\linewidth}\raggedleft
Three attempts, delivered
\end{minipage} & \begin{minipage}[b]{\linewidth}\raggedleft
Three attempts, median
\end{minipage} \\
\midrule\noalign{}
\endhead
\bottomrule\noalign{}
\endlastfoot
GLM-5.3 Flash & 156 & 150 & 96.2\% & 0.7380 & 0.7385 & at least 99.76\%
& 0.7473 \\
GLM-5.3 & 156 & 144 & 92.3\% & 0.7471 & 0.7482 & at least 99.67\% &
0.7564 \\
\end{longtable}

The result also answers a fair question about Studies 1 and 2. When a
benchmark reports that agents do not separate, a reader should ask
whether it can detect anything at all. It can: one planned change moved
the score by seven standard errors on the design that could not rank
three agents. The two arms ran on consecutive days, the Flash runs on 15
and 16 September and the GLM-5.3 runs on 16 and 17 September, through
the same gateway, so the gain includes whatever changed at the endpoint
in between. Scores did not drift within an arm: the first and second
halves of the GLM-5.3 arm average the same to four decimals. And Study
1, which ran both models on 13 September with their runs overlapping in
time, found a gain of similar size for the same three agents, +0.0075
from eight and nine runs (1.7 standard errors). A shift at the endpoint
between days is therefore an unlikely explanation for the gain, though
not one we can exclude.

Table 9 puts the effects on one scale. Each row is a different question;
the last three need very different numbers of runs.

\Needspace{11\baselineskip}

\begin{longtable}[]{@{}
  >{\raggedright\arraybackslash}p{(\linewidth - 6\tabcolsep) * \real{0.7206}}
  >{\raggedleft\arraybackslash}p{(\linewidth - 6\tabcolsep) * \real{0.1176}}
  >{\raggedleft\arraybackslash}p{(\linewidth - 6\tabcolsep) * \real{0.0882}}
  >{\raggedleft\arraybackslash}p{(\linewidth - 6\tabcolsep) * \real{0.0735}}@{}}
\caption{Observed effects against the median within-pairing SD of
GLM-5.3 Flash, 0.0104, and the runs each would need under the
approximation of Section 3.6. The agent rows use the widest gap observed
on each model. The first row compares against a fixed number rather than
a second noisy arm, so the two-arm approximation does not apply to it
and no count is given.}\tabularnewline
\toprule\noalign{}
\begin{minipage}[b]{\linewidth}\raggedright
Question
\end{minipage} & \begin{minipage}[b]{\linewidth}\raggedleft
Effect
\end{minipage} & \begin{minipage}[b]{\linewidth}\raggedleft
In run-to-run SDs
\end{minipage} & \begin{minipage}[b]{\linewidth}\raggedleft
Runs per arm
\end{minipage} \\
\midrule\noalign{}
\endfirsthead
\toprule\noalign{}
\begin{minipage}[b]{\linewidth}\raggedright
Question
\end{minipage} & \begin{minipage}[b]{\linewidth}\raggedleft
Effect
\end{minipage} & \begin{minipage}[b]{\linewidth}\raggedleft
In run-to-run SDs
\end{minipage} & \begin{minipage}[b]{\linewidth}\raggedleft
Runs per arm
\end{minipage} \\
\midrule\noalign{}
\endhead
\bottomrule\noalign{}
\endlastfoot
Did the agent improve on the starting code? & +0.0277 & 2.7 & --- \\
Did the larger model improve the pairing? & +0.0091 & 0.9 & 21 \\
Are two agents different on GLM-5.3? & +0.0067 & 0.6 & 39 \\
Are two agents different on GLM-5.3 Flash? & +0.0039 & 0.4 & 113 \\
\end{longtable}

Improvement over the starting code was much larger than the differences
between configurations. For these observed contrasts, the model change
needs about 21, and telling two agents apart on one model needs roughly
39 to 113.

\subsection{Which agent looks best depends on the
model}\label{which-agent-looks-best-depends-on-the-model}

The upgrade was not worth the same in every agent. pi gained 0.0152,
Hermes 0.0067 and OpenCode 0.0055 (Table 10). pi's gain exceeds
OpenCode's by 0.0097, 3.3 standard errors, and Hermes's by 0.0085, 2.7
standard errors; both remain below 0.05 after Holm adjustment for the
three pairs, and a joint test rejects no interaction (χ² = 12.3 on 2
degrees of freedom, p = 0.002). Hermes and OpenCode cannot be told
apart. The same planned model change, on the same task under the same
budget, was worth more than twice as much in one agent as in either of
the others. This agrees with Lewis (\citeproc{ref-lewis2026same}{2026})
and Fan et al. (\citeproc{ref-fan2026harness}{2026}), who find harness
effects that depend on the model; here the interaction is measured
against the spread of repeated runs. On a later year the pattern keeps
its direction but no longer clears (Section 4.7).

Study 2 shows the same ordering (Figure 3). Moving from GLM-5.3 Flash to
DeepSeek 4.1 Flash raised pi by 0.0095, Hermes by 0.0062 and OpenCode by
0.0003, and the agents' order reverses between the two models: OpenCode
is highest on GLM-5.3 Flash and pi on DeepSeek 4.1 Flash. Two cautions
apply. The pattern was noticed after the data were in, across two models
from different families, so it is a pattern to test rather than a
finding. And the two studies are less independent than they look: both
start from the same GLM-5.3 Flash runs, on which pi trails OpenCode by
0.0039, and that gap counts toward pi's larger gain in both. The part
the studies contribute separately is that pi leads OpenCode on each
stronger model, by 0.0053 on DeepSeek 4.1 Flash and 0.0058 on GLM-5.3,
at 2.4 and 2.6 standard errors (Figure 4).

\begin{figure}[tp]
\centering
\pandocbounded{\includegraphics[keepaspectratio,alt={Study 2's six means read two ways. Left: agents on the axis, one line per model, labelled with how far apart the agents are on it. Right: models on the axis, one line per agent, labelled with its change. Colour is the agent and shape the model. Thick bar: the 95 percent interval of the mean. Pale band: where the middle 80 percent of single runs land. Overlapping bars are not a test; Table 10 gives the tests.}]{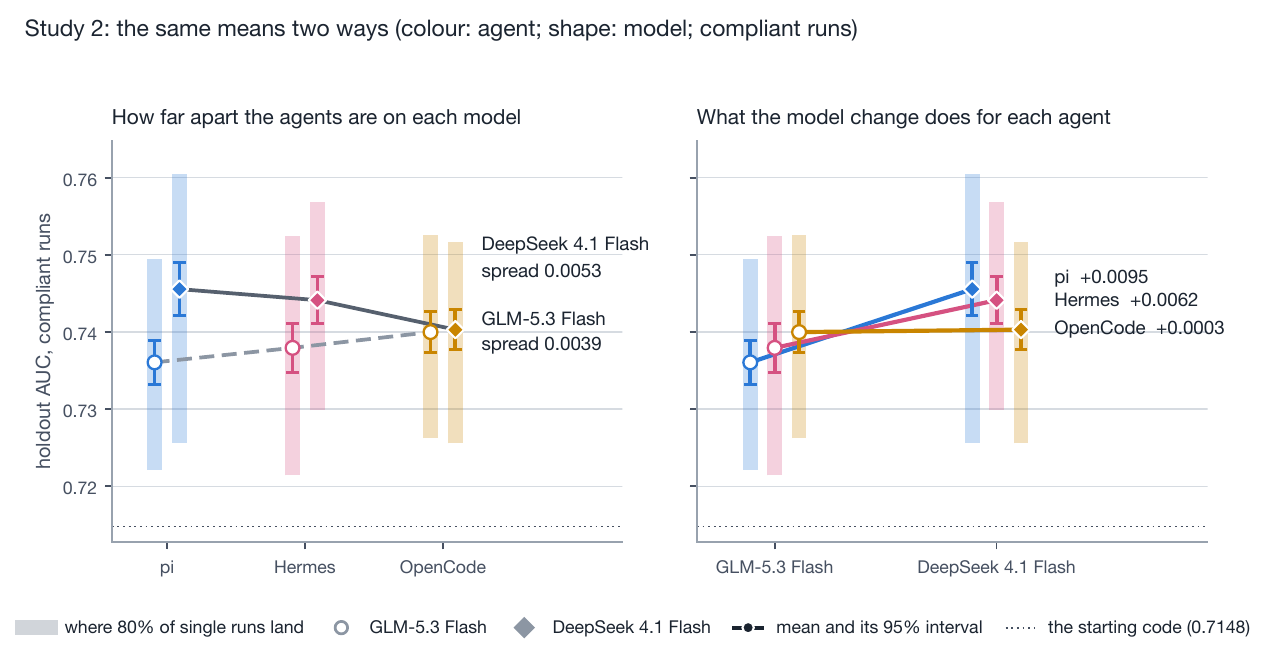}}
\caption{Study 2's six means read two ways. Left: agents on the axis,
one line per model, labelled with how far apart the agents are on it.
Right: models on the axis, one line per agent, labelled with its change.
Colour is the agent and shape the model. Thick bar: the 95 percent
interval of the mean. Pale band: where the middle 80 percent of single
runs land. Overlapping bars are not a test; Table 10 gives the tests.}
\end{figure}

\begin{figure}[tp]
\centering
\pandocbounded{\includegraphics[keepaspectratio,alt={Study 3's means read the same two ways, on the same scale as Figure 3.}]{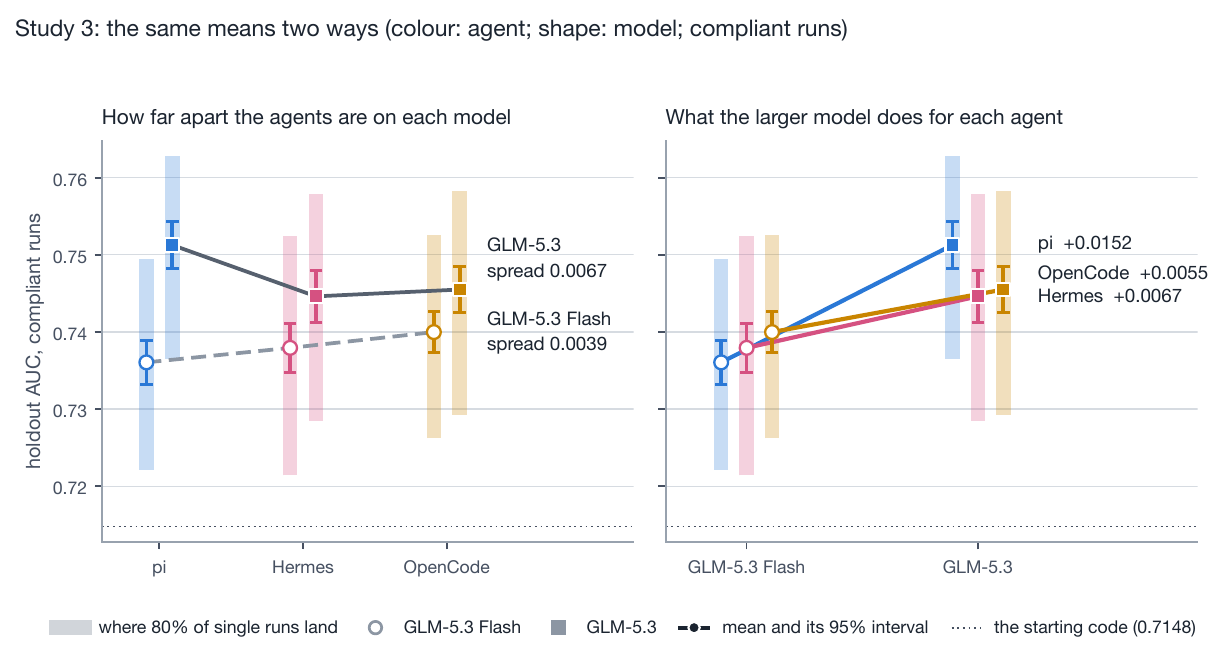}}
\caption{Study 3's means read the same two ways, on the same scale as
Figure 3.}
\end{figure}

\Needspace{13\baselineskip}

\begin{longtable}[]{@{}
  >{\raggedright\arraybackslash}p{(\linewidth - 12\tabcolsep) * \real{0.2577}}
  >{\raggedright\arraybackslash}p{(\linewidth - 12\tabcolsep) * \real{0.2268}}
  >{\raggedleft\arraybackslash}p{(\linewidth - 12\tabcolsep) * \real{0.0825}}
  >{\raggedleft\arraybackslash}p{(\linewidth - 12\tabcolsep) * \real{0.0722}}
  >{\raggedleft\arraybackslash}p{(\linewidth - 12\tabcolsep) * \real{0.0619}}
  >{\raggedright\arraybackslash}p{(\linewidth - 12\tabcolsep) * \real{0.2371}}
  >{\raggedleft\arraybackslash}p{(\linewidth - 12\tabcolsep) * \real{0.0619}}@{}}
\caption{Interaction contrasts: the change in one agent minus the change
in another when the model changes from GLM-5.3 Flash, compliant runs.
All three pairs are shown for each study; the choice of pair is
exploratory in both. Joint tests of no interaction: χ² = 10.2, p = 0.006
(Study 2); χ² = 12.3, p = 0.002 (Study 3).}\tabularnewline
\toprule\noalign{}
\begin{minipage}[b]{\linewidth}\raggedright
Study: model change
\end{minipage} & \begin{minipage}[b]{\linewidth}\raggedright
Contrast
\end{minipage} & \begin{minipage}[b]{\linewidth}\raggedleft
Estimate
\end{minipage} & \begin{minipage}[b]{\linewidth}\raggedleft
SE
\end{minipage} & \begin{minipage}[b]{\linewidth}\raggedleft
In SEs
\end{minipage} & \begin{minipage}[b]{\linewidth}\raggedright
Pointwise 95\% interval
\end{minipage} & \begin{minipage}[b]{\linewidth}\raggedleft
Holm p
\end{minipage} \\
\midrule\noalign{}
\endfirsthead
\toprule\noalign{}
\begin{minipage}[b]{\linewidth}\raggedright
Study: model change
\end{minipage} & \begin{minipage}[b]{\linewidth}\raggedright
Contrast
\end{minipage} & \begin{minipage}[b]{\linewidth}\raggedleft
Estimate
\end{minipage} & \begin{minipage}[b]{\linewidth}\raggedleft
SE
\end{minipage} & \begin{minipage}[b]{\linewidth}\raggedleft
In SEs
\end{minipage} & \begin{minipage}[b]{\linewidth}\raggedright
Pointwise 95\% interval
\end{minipage} & \begin{minipage}[b]{\linewidth}\raggedleft
Holm p
\end{minipage} \\
\midrule\noalign{}
\endhead
\bottomrule\noalign{}
\endlastfoot
2: to DeepSeek 4.1 Flash & pi vs Hermes & +0.0033 & 0.0032 & 1.0 &
−0.0029 to +0.0094 & 0.30 \\
& pi vs OpenCode & +0.0092 & 0.0030 & 3.1 & +0.0034 to +0.0150 &
0.006 \\
& Hermes vs OpenCode & +0.0059 & 0.0030 & 2.0 & +0.0002 to +0.0116 &
0.095 \\
3: to GLM-5.3 & pi vs Hermes & +0.0085 & 0.0032 & 2.7 & +0.0024 to
+0.0148 & 0.015 \\
& pi vs OpenCode & +0.0097 & 0.0029 & 3.3 & +0.0040 to +0.0155 &
0.003 \\
& Hermes vs OpenCode & +0.0012 & 0.0031 & 0.4 & −0.0049 to +0.0072 &
0.71 \\
\end{longtable}

The pale bands in Figures 3 and 4 separate the two kinds of spread this
paper is about. Every Study 2 mean lies inside every pairing's middle 80
percent of runs, and even pi's two Study 3 bands overlap across their
middle, although pi's gain is 7.1 standard errors from zero in the
means. The means are precise; a comparison of single runs is unreliable.

\subsection{A later year keeps a third of the
gain}\label{a-later-year-keeps-a-third-of-the-gain}

Every score so far comes from a holdout drawn from 2006, the year of the
evaluation set the agents tuned on. To see how much of what they
delivered carries forward, we retrained all 464 scored programs of
Studies 2 and 3 exactly as the scorer does and applied each to one
million flights from 2007, prepared the same way and used by no earlier
analysis (Table 11). The same fits reproduced the recorded 2006 scores
(median difference 0.0002 among compliant runs), and the starting code
scores 0.7175 on 2007, slightly above its 0.7148 on 2006, so it did not
score worse on the later-year data.

\Needspace{14\baselineskip}

\begin{longtable}[]{@{}
  >{\raggedright\arraybackslash}p{(\linewidth - 4\tabcolsep) * \real{0.6235}}
  >{\raggedleft\arraybackslash}p{(\linewidth - 4\tabcolsep) * \real{0.1882}}
  >{\raggedleft\arraybackslash}p{(\linewidth - 4\tabcolsep) * \real{0.1882}}@{}}
\caption{The same delivered programs on the 2006 holdout and on 2007
flights. The policy is that of Table 6, chosen on the evaluation set as
before; its 2007 gain has a 95 percent interval of +0.0010 to
+0.0032.}\tabularnewline
\toprule\noalign{}
\begin{minipage}[b]{\linewidth}\raggedright
\end{minipage} & \begin{minipage}[b]{\linewidth}\raggedleft
2006 holdout
\end{minipage} & \begin{minipage}[b]{\linewidth}\raggedleft
2007 flights
\end{minipage} \\
\midrule\noalign{}
\endfirsthead
\toprule\noalign{}
\begin{minipage}[b]{\linewidth}\raggedright
\end{minipage} & \begin{minipage}[b]{\linewidth}\raggedleft
2006 holdout
\end{minipage} & \begin{minipage}[b]{\linewidth}\raggedleft
2007 flights
\end{minipage} \\
\midrule\noalign{}
\endhead
\bottomrule\noalign{}
\endlastfoot
Gain over the starting code, 446 compliant runs & +0.0280 & +0.0092 \\
Compliant runs below the starting code & 0 & 33 \\
Spread of the six Study 2 pairing means & 0.0095 & 0.0021 \\
Median run-to-run SD, Study 2 & 0.0107 & 0.0060 \\
Best of three attempts over one & +0.0081 & +0.0022 \\
Larger model's gain & +0.0091 (7.1 SE) & +0.0027 (3.8 SE) \\
pi's gain minus OpenCode's & +0.0097 (3.3 SE) & +0.0023 (1.3 SE) \\
Rule-breakers among the ten highest, Studies 2 and 3 & 7 and 6 & 7 and
6 \\
\end{longtable}

A third of the agents' gain survives. The rest was specific to 2006, the
year both the evaluation set and the holdout come from, and a run's two
scores are only loosely related: inside a pairing their rank correlation
is +0.38. Every finding keeps its direction on 2007 but shrinks. Runs of
one pairing still vary more than the pairings differ, now 0.0060 against
0.0021. The rule-breaking runs still hold the top of the ranking, and
those that trained on the evaluation labels stay well above the
compliant runs, consistent with a real gain from more recent data. Three
attempts still buy a better artifact, by about a quarter as much. The
larger model's gain remains detectable but falls to 0.44 run-to-run SDs,
about 82 runs per arm under the approximation of Section 3.6, and pi's
larger share of it no longer separates from the other agents'.

The sizes in this paper therefore describe the year the agents tuned on.
That is what a team measures when it validates on the period it selected
on, and it is why confirming a chosen model on later, untouched data
matters more than any single number here. We did not test why the gain
shrinks.

\subsection{Budget use predicts score inside a
pairing}\label{budget-use-predicts-score-inside-a-pairing}

Runs that spent more of their measured compute scored higher (Figure 5).
In Study 1 the rank correlation between budget use and score is +0.59
over 103 runs, where it could mean that some agents simply work harder
than others. Study 2 holds the agent and model fixed: within pairings
the correlation is +0.55 over 312 runs (95 percent interval +0.45 to
+0.63), the same over compliant runs, and positive in every pairing,
from +0.38 to +0.70. The association is correlational: it does not show
that forcing a run to use more CPU would improve it, and an agent whose
search is going well may simply continue.

\begin{figure}
\centering
\pandocbounded{\includegraphics[keepaspectratio,alt={Budget use against score inside each pairing, compliant runs only. Colour is the agent and shape the model. Dotted line: the starting code.}]{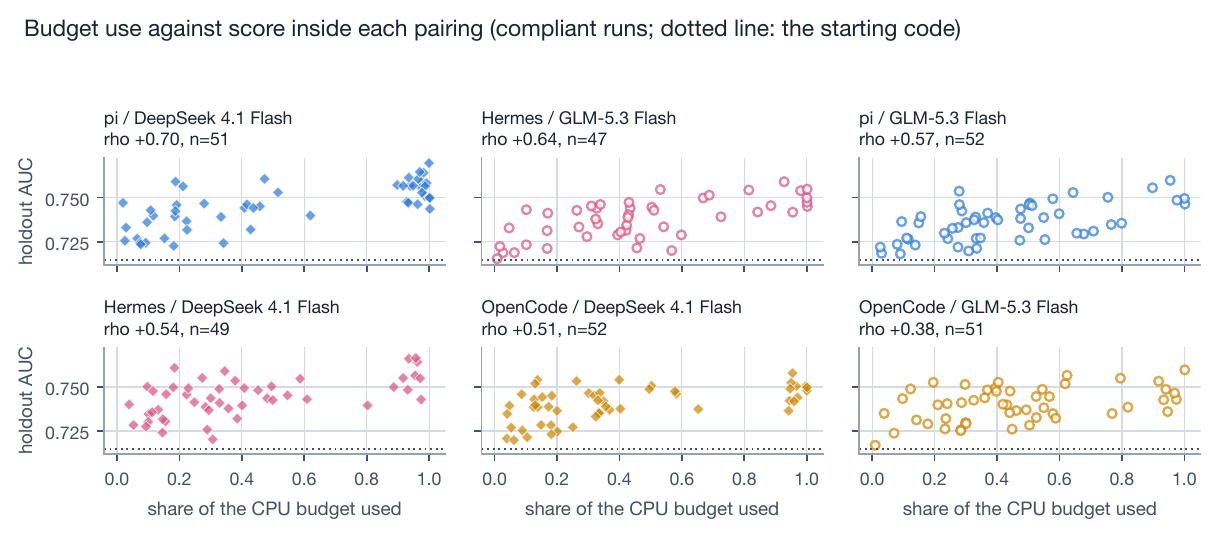}}
\caption{Budget use against score inside each pairing, compliant runs
only. Colour is the agent and shape the model. Dotted line: the starting
code.}
\end{figure}

Low budget use is not a sign of giving up. Every one of the 95 runs that
used under a quarter of the budget ran all 40 counted experiments; they
used cheaper methods. The association weakens among high-budget runs
(+0.43 above a quarter of the budget, +0.29 above half, +0.12 above
three quarters with an interval spanning zero), though narrowing the
sample also narrows the predictor, so this is not by itself evidence of
diminishing returns (Table 16).

\subsection{Cost follows the pairing, mostly through the
cache}\label{cost-follows-the-pairing-mostly-through-the-cache}

In Study 1, on DeepSeek V4 Flash through Ollama Cloud, the same job cost
\$0.08 a run through pi and \$1.86 through Claude Code at rate-card
prices (Table 12). This 23-fold figure is a rate-card projection for one
observed comparison, not a general multiplier for either agent. The
endpoint served 92 to 98 percent of pi's and Codex's input from its
prompt cache, and 2 to 11 percent of Claude Code's. Repricing Claude
Code's own token volume at pi's cache share gives \$0.20 a run: as an
accounting counterfactual, the cache accounts for about ninefold of the
gap and Claude Code's larger token volume for the rest. The
counterfactual does not explain why the cache failed.

\Needspace{11\baselineskip}

\begin{longtable}[]{@{}
  >{\raggedright\arraybackslash}p{(\linewidth - 8\tabcolsep) * \real{0.2059}}
  >{\raggedleft\arraybackslash}p{(\linewidth - 8\tabcolsep) * \real{0.2059}}
  >{\raggedleft\arraybackslash}p{(\linewidth - 8\tabcolsep) * \real{0.2059}}
  >{\raggedleft\arraybackslash}p{(\linewidth - 8\tabcolsep) * \real{0.2059}}
  >{\raggedleft\arraybackslash}p{(\linewidth - 8\tabcolsep) * \real{0.1765}}@{}}
\caption{Study 1, cached share of input and rate-card cost per run,
DeepSeek V4 Flash, Ollama Cloud.}\tabularnewline
\toprule\noalign{}
\begin{minipage}[b]{\linewidth}\raggedright
Agent
\end{minipage} & \begin{minipage}[b]{\linewidth}\raggedleft
Run 1
\end{minipage} & \begin{minipage}[b]{\linewidth}\raggedleft
Run 2
\end{minipage} & \begin{minipage}[b]{\linewidth}\raggedleft
Run 3
\end{minipage} & \begin{minipage}[b]{\linewidth}\raggedleft
Mean cost
\end{minipage} \\
\midrule\noalign{}
\endfirsthead
\toprule\noalign{}
\begin{minipage}[b]{\linewidth}\raggedright
Agent
\end{minipage} & \begin{minipage}[b]{\linewidth}\raggedleft
Run 1
\end{minipage} & \begin{minipage}[b]{\linewidth}\raggedleft
Run 2
\end{minipage} & \begin{minipage}[b]{\linewidth}\raggedleft
Run 3
\end{minipage} & \begin{minipage}[b]{\linewidth}\raggedleft
Mean cost
\end{minipage} \\
\midrule\noalign{}
\endhead
\bottomrule\noalign{}
\endlastfoot
pi & 98\%, \$0.08 & 98\%, \$0.11 & 97\%, \$0.05 & \$0.08 \\
OpenClaw & no record & 98\%, \$0.11 & 97\%, \$0.07 & \$0.09 \\
Codex & 96\%, \$0.14 & 96\%, \$0.19 & 92\%, \$0.13 & \$0.15 \\
OpenCode & 80\%, \$0.24 & 64\%, \$0.19 & 66\%, \$0.49 & \$0.31 \\
Hermes & 56\%, \$1.09 & 89\%, \$0.47 & 74\%, \$0.24 & \$0.60 \\
Claude Code & 11\%, \$1.41 & 9\%, \$2.95 & 2\%, \$1.21 & \$1.86 \\
\end{longtable}

The cache failure belonged to the pairing, not to either half. Claude
Code cached 96 to 97 percent of its input on DeepSeek V4.1 Flash and 73
to 84 percent on Nemotron Super, and every agent on those two models
cached 70 to 99 percent. It failed again on one more model, DeepSeek V4
Pro, whose row the quota stopped after eight runs: Claude Code cached 4
and 11 percent while the other five agents cached 83 to 99. The quota
cut off both Claude Code runs, but also both of pi's, which still cached
91 and 98 percent. Claude Code's session transcripts show the failure
request by request. On V4 Flash none of its 338 requests read back more
than half its prompt from the cache, and no request read back more than
17,024 tokens, as prompts grew to 164,044; on V4.1 Flash the same agent
read back more than half its prompt on 296 of 303 requests, up to
146,304 tokens. Context size does not explain it.

Lawrence (\citeproc{ref-lawrence2026harness}{2026}) observed the same
agent's cache share collapse through another gateway and traced it to
the gateway's translation of Claude Code's Anthropic-style request
format. Claude Code reached Ollama Cloud through the same kind of
endpoint, so our case is consistent with his conclusion that cache share
belongs to the whole path from agent through gateway to provider. It is
not consistent with a translation that fails for every model: through
the same endpoint, Claude Code cached normally on V4.1 Flash and
Nemotron Super. Whatever failed depends on the model behind the endpoint
as well. We have not tested the mechanism. Cost also moved between
identical runs: Hermes cost \$0.24, \$0.47 and \$1.09 for the same work,
a 4.6-fold swing, and across the 25 Study 1 pairings with three complete
costs the median swing was 2.2-fold.

Table 13 sets the nine configurations of Studies 2 and 3 side by side:
their token use, list-price cost, yield, quality and the three-attempt
policy. Within a model, the costliest agent spent 1.2 to 2.1 times what
the cheapest did per run, and the order by cost did not follow the order
by quality: Hermes read the most input on every model without delivering
the best artifacts. Moving from GLM-5.3 Flash to GLM-5.3 raised
list-price cost about fifteenfold for 0.0091 AUC. These costs apply no
cache discount, which Study 1 shows can reorder agents, and exclude
reasoning tokens; over Study 2's runs, reasoning was about a third of
all generated tokens, 8.7 of 25.1 million.

\Needspace{16\baselineskip}

\begin{longtable}[]{@{}
  >{\raggedright\arraybackslash}p{(\linewidth - 12\tabcolsep) * \real{0.4167}}
  >{\raggedleft\arraybackslash}p{(\linewidth - 12\tabcolsep) * \real{0.0972}}
  >{\raggedleft\arraybackslash}p{(\linewidth - 12\tabcolsep) * \real{0.0972}}
  >{\raggedleft\arraybackslash}p{(\linewidth - 12\tabcolsep) * \real{0.0972}}
  >{\raggedleft\arraybackslash}p{(\linewidth - 12\tabcolsep) * \real{0.0972}}
  >{\raggedleft\arraybackslash}p{(\linewidth - 12\tabcolsep) * \real{0.0972}}
  >{\raggedleft\arraybackslash}p{(\linewidth - 12\tabcolsep) * \real{0.0972}}@{}}
\caption{The observed configurations of Studies 2 and 3, per run. Costs
apply OpenRouter list prices of 16 September 2026 with no cache discount
and exclude reasoning tokens; they compare configurations and are not
bills. Yield is the share of 52 runs that were compliant and scored.
Appendix D reprices Study 2's token ledger at other models'
prices.}\tabularnewline
\toprule\noalign{}
\begin{minipage}[b]{\linewidth}\raggedright
Agent, model
\end{minipage} & \begin{minipage}[b]{\linewidth}\raggedleft
Input tokens per run
\end{minipage} & \begin{minipage}[b]{\linewidth}\raggedleft
Output tokens per run
\end{minipage} & \begin{minipage}[b]{\linewidth}\raggedleft
List-price cost per run
\end{minipage} & \begin{minipage}[b]{\linewidth}\raggedleft
Yield
\end{minipage} & \begin{minipage}[b]{\linewidth}\raggedleft
Compliant mean
\end{minipage} & \begin{minipage}[b]{\linewidth}\raggedleft
Best of 3, median
\end{minipage} \\
\midrule\noalign{}
\endfirsthead
\toprule\noalign{}
\begin{minipage}[b]{\linewidth}\raggedright
Agent, model
\end{minipage} & \begin{minipage}[b]{\linewidth}\raggedleft
Input tokens per run
\end{minipage} & \begin{minipage}[b]{\linewidth}\raggedleft
Output tokens per run
\end{minipage} & \begin{minipage}[b]{\linewidth}\raggedleft
List-price cost per run
\end{minipage} & \begin{minipage}[b]{\linewidth}\raggedleft
Yield
\end{minipage} & \begin{minipage}[b]{\linewidth}\raggedleft
Compliant mean
\end{minipage} & \begin{minipage}[b]{\linewidth}\raggedleft
Best of 3, median
\end{minipage} \\
\midrule\noalign{}
\endhead
\bottomrule\noalign{}
\endlastfoot
pi, GLM-5.3 Flash & 4.34M & 58k & \$0.45 & 100\% & 0.7361 & 0.7460 \\
Hermes, GLM-5.3 Flash & 8.12M & 68k & \$0.83 & 90.4\% & 0.7379 &
0.7461 \\
OpenCode, GLM-5.3 Flash & 5.70M & 32k & \$0.58 & 98.1\% & 0.7400 &
0.7486 \\
pi, DeepSeek 4.1 Flash & 3.93M & 63k & \$0.63 & 98.1\% & 0.7456 &
0.7567 \\
Hermes, DeepSeek 4.1 Flash & 8.65M & 69k & \$1.34 & 94.2\% & 0.7441 &
0.7535 \\
OpenCode, DeepSeek 4.1 Flash & 5.07M & 26k & \$0.78 & 100\% & 0.7403 &
0.7488 \\
pi, GLM-5.3 & 5.87M & 91k & \$8.61 & 90.4\% & 0.7513 & 0.7594 \\
Hermes, GLM-5.3 & 7.43M & 62k & \$10.67 & 90.4\% & 0.7446 & 0.7537 \\
OpenCode, GLM-5.3 & 6.40M & 38k & \$9.13 & 96.2\% & 0.7455 & 0.7563 \\
\end{longtable}

\section{Discussion}\label{discussion}

\subsection{What this means for evaluating
agents}\label{what-this-means-for-evaluating-agents}

The first implication is the unit. Neither the agent nor the model
predicted how a combination would score or what it would cost: the
ranking of agents reversed between models, a planned model upgrade was
worth more than twice as much in one agent as in the others, and a
prompt cache that worked for an agent on one model failed on another.
This matches the conclusion Lewis (\citeproc{ref-lewis2026same}{2026})
draws from harness interventions. An evaluation should name the pairing
it tested.

The second is the number of runs. Table 9 shows how much the answer
depends on the question. The gain over the starting code was far larger
than any other contrast, while detecting this model upgrade needs about
21 runs of each and telling apart the agents observed here on one model
roughly 39 to 113. Few repetitions can reveal large operational
differences, as Study 1's cache failure shows, but they were
insufficient to characterise within-task variability or to resolve
reliably the small agent differences observed here. That is ordinary
power analysis (\citeproc{ref-miller2024errorbars}{Miller 2024}); the
measured run-to-run spread is what sets the numbers. The sizes
themselves are those of the year the agents tuned on: on a later year
the gains shrank by two thirds (Section 4.7). Designs with many tasks
and few attempts per task answer a different question, how a pairing
does across a task distribution, and this study does not speak to their
adequacy for it.

The third is the difference between choosing and ranking. A few attempts
buy a better artifact cheaply: at the token prices of this study,
roughly \$2 for three attempts and \$8 for ten, before the audit work.
Ranking two vendors on this task would take tens of runs of each. And
because the highest scores are disproportionately noncompliant, every
selection policy needs an audit before the ranking, not after it.

\subsection{What a difference is
worth}\label{what-a-difference-is-worth}

An AUC gap should not be converted into money. AUC summarises ranking
across every threshold, so it does not say what changes at the one
threshold a team uses. Precision-recall behaviour, unlike ROC, changes
with the ratio of positives to negatives
(\citeproc{ref-saito2015precision}{Saito and Rehmsmeier 2015}), and
decision curve analysis evaluates a model by its net benefit at the
thresholds a decision-maker might use, each encoding the relative harm
of false positives and false negatives
(\citeproc{ref-vickers2006decision}{Vickers and Elkin 2006}). Score the
candidates on held-back data, apply the decision rule you run, and count
the outcomes at your own prevalence and costs. Table 14 does this for
two of our models, the best compliant run of a pairing and a run near
its median (0.7695 and 0.7471 AUC; pi on DeepSeek 4.1 Flash, whose
compliant median is 0.7462), for a team that sees one million cases a
month, one in a hundred of them positive, with a false alarm costing
\$10 and a miss \$100, assuming only the prevalence differs from the
benchmark.

\Needspace{9\baselineskip}

\begin{longtable}[]{@{}
  >{\raggedright\arraybackslash}p{(\linewidth - 6\tabcolsep) * \real{0.3143}}
  >{\raggedleft\arraybackslash}p{(\linewidth - 6\tabcolsep) * \real{0.2286}}
  >{\raggedleft\arraybackslash}p{(\linewidth - 6\tabcolsep) * \real{0.2286}}
  >{\raggedleft\arraybackslash}p{(\linewidth - 6\tabcolsep) * \real{0.2286}}@{}}
\caption{Illustrative, not a forecast. Both models were selected and
scored on the same holdout, so these are not validated savings. The
better model by AUC is worse at the tightest capacity.}\tabularnewline
\toprule\noalign{}
\begin{minipage}[b]{\linewidth}\raggedright
Cases reviewed each month
\end{minipage} & \begin{minipage}[b]{\linewidth}\raggedleft
Caught by the 0.7695 model
\end{minipage} & \begin{minipage}[b]{\linewidth}\raggedleft
Caught by the 0.7471 model
\end{minipage} & \begin{minipage}[b]{\linewidth}\raggedleft
Difference
\end{minipage} \\
\midrule\noalign{}
\endfirsthead
\toprule\noalign{}
\begin{minipage}[b]{\linewidth}\raggedright
Cases reviewed each month
\end{minipage} & \begin{minipage}[b]{\linewidth}\raggedleft
Caught by the 0.7695 model
\end{minipage} & \begin{minipage}[b]{\linewidth}\raggedleft
Caught by the 0.7471 model
\end{minipage} & \begin{minipage}[b]{\linewidth}\raggedleft
Difference
\end{minipage} \\
\midrule\noalign{}
\endhead
\bottomrule\noalign{}
\endlastfoot
5,000 & 562 & 603 & −\$4,501 \\
20,000 & 1,351 & 1,287 & \$7,031 \\
100,000 & 3,714 & 3,462 & \$27,665 \\
\end{longtable}

The crossing is the point. Rank candidates on the objective you will
use, on a selection set, and measure the chosen one on untouched data.
Average precision needs the same care: for fixed score distributions
within each class, precision, and with it the precision-recall curve and
AP, changes with prevalence (\citeproc{ref-saito2015precision}{Saito and
Rehmsmeier 2015}). Our holdout was built with the classes in equal
numbers while real delays are far rarer, so our AP values compare runs
fairly here and are not the precision a deployment would see.

\subsection{Relation to prior work}\label{relation-to-prior-work}

Run-to-run variation, harness effects that depend on the model,
best-of-k evaluation and rule-breaking by agents are each established
(Section 2). Bjarnason et al.
(\citeproc{ref-bjarnason2026randomness}{2026}) already showed that
single runs of coding agents mislead and that the number of runs should
come from a power analysis; our run counts agree with theirs in kind.
Chen et al. (\citeproc{ref-chen2026publicscore}{2026}) already showed
that agents exploit labelled evaluation data, and that the exploitation
need not help on hidden data. What this design adds is their joint
measurement on one task under one protocol, on a continuous measure of
the delivered artifact, with enough repetitions per pairing to express a
planned model upgrade, an agent-by-model interaction and the value of a
repeat-and-select policy in units of the run-to-run spread. It separates
two uses of repeated attempts that are easily confused: selecting an
artifact, which a few attempts do well once compliance is checked first,
and ranking configurations, which takes tens of runs. Compliance is
reported as an outcome beside quality rather than as data cleaning, and
the later year shows how much of each gain the tuning year overstated.

\subsection{Limitations}\label{limitations}

\textbf{One task.} Everything here comes from one tabular prediction
task with one dataset. The phenomena are likely to appear elsewhere;
their sizes may not. Benchmarks with many tasks and pass/fail outcomes
estimate how results generalise across tasks; ours estimates variation
within one configuration and small effects on one task. Neither replaces
the other.

\textbf{Exploratory interaction.} Study 2's agent-by-model pattern was
noticed after the data were in, and which agent pair to compare in Study
3 was not planned; the Holm adjustment and joint test cover the three
pairs within each study, not the choice to look. The two studies share
their GLM-5.3 Flash runs, so they are not independent confirmations. A
replication specified in advance, on another model family, would test
whether the pattern generalises.

\textbf{Hosted endpoints change.} We measured systems as deployed,
through hosted endpoints whose builds, load and routing we did not pin
and could not have held constant. That variation is part of what we
report, as it would be for anyone using the same services. It matters
most for Study 3, whose arms ran on consecutive days, so its gain
includes any change at the endpoint between them; the same-day
comparison in Study 1 points the same way.

\textbf{The year the agents tuned on.} The evaluation set and the
holdout both come from 2006, so the headline sizes include what is
specific to that year. Section 4.7 shows how much of each survives on
2007 flights.

\textbf{A retrospective policy analysis.} The repeat-and-select policy
was specified after the runs had been scored. The holdout that scores it
was hidden from the agents, but it had informed our earlier analyses, so
the policy result has the standing of a retrospective evaluation rather
than a locked, prospective one.

\textbf{The task's features.} The departure-time field records when the
flight actually left. That does not affect the comparisons, but the AUCs
are not those of a forecast made before departure.

\textbf{Selected agents and endpoints.} Study 2's three agents were the
highest-scoring in Study 1's LunaRoute rows. The endpoints are two
hosted services, and the same model name can behave differently across
providers.

\textbf{What the meters see.} The compute meter excludes the model's
generation and in-context reasoning. Costs are rate-card projections,
not invoices, and the reasoning setting was left at each endpoint's
default, which for the GLM models is the maximum. Early stopping on the
evaluation set was permitted, which makes each run's reported evaluation
score optimistic.

\textbf{Compliance is decided by traces and reading.} The traces look
for the two violations we anticipated or found, and every hit was read,
but a violation of a kind no trace looks for would pass. One compliant
run is a judgment call (Section 3.5), and it does not move any result.

\subsection{Future work}\label{future-work}

A replication of the agent-by-model interaction specified in advance, on
a different model family, would show whether pi's larger gains
generalise. Sending byte-identical prompts to one endpoint in both
request formats, the Anthropic-style one Claude Code uses and the
OpenAI-style one the other agents use, would test whether the mechanism
Lawrence (\citeproc{ref-lawrence2026harness}{2026}) found explains our
cache failure, and why it appears on some models only. Re-scoring a
sample of delivered artifacts many times would bound the retraining
component of the spread more tightly than our spot checks. A broader
question is how context management interacts with caching: strategies
that rewrite or summarise earlier context, which Fan et al.
(\citeproc{ref-fan2026harness}{2026}) find the most efficient when
tokens are priced at list rates, also change the prompt prefix a cache
depends on, and could rank differently once cached input is priced as it
is billed.

\section{Conclusion}\label{conclusion}

For this fixed ML-engineering task and these deployed configurations,
repeated attempts revealed substantial variation in outcome, compliance
failures distorted the upper tail, and a small number of attempts, with
compliance checked first, supported useful artifact selection even where
precise comparisons between configurations required substantially more
evidence. A planned model upgrade was worth about one run of noise, and
a different amount in each agent; on a later year the delivered models
kept a third of their gain; cost depended on whether the prompt cache
worked for that pairing. Few repetitions can reveal large operational
differences, but they were insufficient to characterise the variation
within a task or to resolve the small agent differences observed here.
For teams deploying agents on tasks like this one:

\begin{itemize}
\tightlist
\item
  \textbf{Evaluate the pairing, not the parts,} and record whether
  attempts started and completed, not only how finished ones scored.
\item
  \textbf{Attempt the job several times.} The best compliant artifact
  among three attempts had a median 0.008 AUC above a single attempt;
  ten attempts added little beyond three.
\item
  \textbf{Reject first, then rank.} Check every attempt against your
  rules before comparing scores, and where possible make violations
  impossible rather than detectable.
\item
  \textbf{Rank on your objective, then confirm on later, untouched
  data,} and read the leader's code. Here a later year kept a third of
  the gain the tuning year showed.
\item
  \textbf{Instrument what nobody looks at:} the cached share of input
  and the compute used on every run.
\item
  \textbf{Price the winner with its predictions,} at your prevalence and
  costs, and price the failures too.
\item
  \textbf{Test again when anything changes:} the model, the agent
  version, the endpoint, the cache or the price.
\end{itemize}

\section*{References}\label{references}
\addcontentsline{toc}{section}{References}

\protect\phantomsection\label{refs}

\clearpage
\appendix

\section{Study 1 in detail}\label{study-1-in-detail}

\Needspace{11\baselineskip}

\begin{longtable}[]{@{}lrrrr@{}}
\caption{Study 1 run flow: 116 runs, 113 of which delivered an artifact.
Three Codex runs never started because the gateway rejected the request
format, leaving the starting code in place; that failure is a property
of the pairing and stays in the table. The V4 Pro row was stopped at 85
percent of a weekly quota, four of its runs cut off mid-run, and appears
only in the cost analysis.}\tabularnewline
\toprule\noalign{}
Model rows & Scheduled & Started & Delivered an artifact & Compliant \\
\midrule\noalign{}
\endfirsthead
\toprule\noalign{}
Model rows & Scheduled & Started & Delivered an artifact & Compliant \\
\midrule\noalign{}
\endhead
\bottomrule\noalign{}
\endlastfoot
Five complete rows & 90 & 90 & 90 & 88 \\
DeepSeek 4.1 Flash, LunaRoute & 18 & 15 & 15 & 15 \\
DeepSeek V4 Pro, Ollama (partial) & 8 & 8 & 8 & not assessed \\
\end{longtable}

\begin{figure}
\centering
\pandocbounded{\includegraphics[keepaspectratio,alt={Holdout AUC, one panel per model row of Study 1. Per agent: the mean of three runs (large mark), the runs (small marks) and their range (bar). pi, Hermes and OpenCode keep their colours; the other three agents are grey.}]{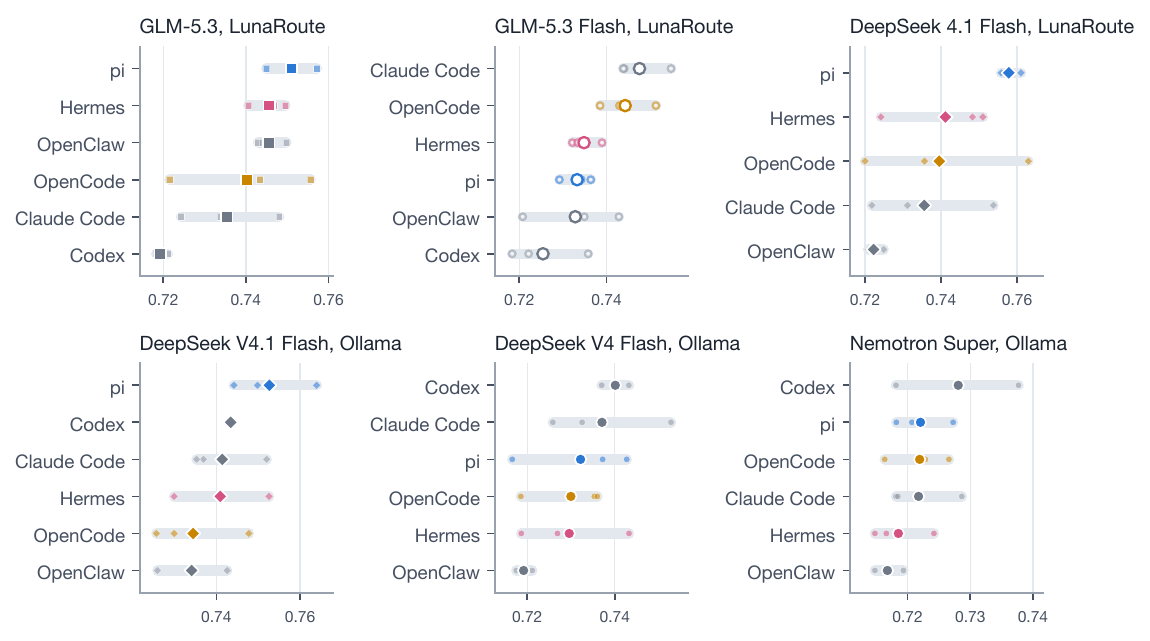}}
\caption{Holdout AUC, one panel per model row of Study 1. Per agent: the
mean of three runs (large mark), the runs (small marks) and their range
(bar). pi, Hermes and OpenCode keep their colours; the other three
agents are grey.}
\end{figure}

Study 1's quality tables exclude five of the 108 runs on its complete
rows: the three that never started, one that concatenated the evaluation
rows into its training data, and one whose delivered code computed count
features on the data being scored, leaving 103. For two agents chosen in
advance, the three-run design separates a difference of about 0.02 AUC.

\section{The compute budget in
detail}\label{the-compute-budget-in-detail}

The meter covers every Python process started inside the container:
experiments run through the provided script, the agent's own scripts,
one-line commands, data loading and feature engineering. A start-up hook
records each process on exit, and every model fit is logged separately
with its row count and duration. The budget was calibrated by re-running
earlier delivered code on the same hardware. Counting fits instead of
CPU seconds predicts score far more weakly: +0.25 over Study 1's runs
and +0.09 within Study 2's pairings. The number of fits says little; the
compute they consume says more.

\section{Statistical details}\label{statistical-details}

\subsection{Compute against score}\label{compute-against-score}

\Needspace{12\baselineskip}

\begin{longtable}[]{@{}
  >{\raggedright\arraybackslash}p{(\linewidth - 8\tabcolsep) * \real{0.4805}}
  >{\raggedleft\arraybackslash}p{(\linewidth - 8\tabcolsep) * \real{0.0779}}
  >{\raggedleft\arraybackslash}p{(\linewidth - 8\tabcolsep) * \real{0.0909}}
  >{\raggedright\arraybackslash}p{(\linewidth - 8\tabcolsep) * \real{0.2078}}
  >{\raggedleft\arraybackslash}p{(\linewidth - 8\tabcolsep) * \real{0.1429}}@{}}
\caption{Study 2, rank correlation between budget use and score, with
runs ranked inside their own pairing and then pooled. Intervals are
bootstrap over runs with 2,000 resamples; p-values are two-sided, from
permutation within pairings. Study 1: +0.59 over 103 runs, interval
+0.45 to +0.71.}\tabularnewline
\toprule\noalign{}
\begin{minipage}[b]{\linewidth}\raggedright
Subset
\end{minipage} & \begin{minipage}[b]{\linewidth}\raggedleft
Runs
\end{minipage} & \begin{minipage}[b]{\linewidth}\raggedleft
Rank correlation
\end{minipage} & \begin{minipage}[b]{\linewidth}\raggedright
95\% interval
\end{minipage} & \begin{minipage}[b]{\linewidth}\raggedleft
p
\end{minipage} \\
\midrule\noalign{}
\endfirsthead
\toprule\noalign{}
\begin{minipage}[b]{\linewidth}\raggedright
Subset
\end{minipage} & \begin{minipage}[b]{\linewidth}\raggedleft
Runs
\end{minipage} & \begin{minipage}[b]{\linewidth}\raggedleft
Rank correlation
\end{minipage} & \begin{minipage}[b]{\linewidth}\raggedright
95\% interval
\end{minipage} & \begin{minipage}[b]{\linewidth}\raggedleft
p
\end{minipage} \\
\midrule\noalign{}
\endhead
\bottomrule\noalign{}
\endlastfoot
All runs & 312 & +0.55 & +0.45 to +0.63 & below 0.001 \\
Compliant runs & 302 & +0.55 & +0.46 to +0.63 & below 0.001 \\
Compliant, above a quarter of budget & 209 & +0.43 & +0.29 to +0.54 &
below 0.001 \\
Compliant, above half & 110 & +0.29 & +0.10 to +0.45 & 0.001 \\
Compliant, above three quarters & 70 & +0.12 & −0.15 to +0.41 & 0.342 \\
\end{longtable}

\subsection{Best-of-k for each
pairing}\label{best-of-k-for-each-pairing}

\Needspace{12\baselineskip}

\begin{longtable}[]{@{}
  >{\raggedright\arraybackslash}p{(\linewidth - 12\tabcolsep) * \real{0.3846}}
  >{\raggedleft\arraybackslash}p{(\linewidth - 12\tabcolsep) * \real{0.1026}}
  >{\raggedleft\arraybackslash}p{(\linewidth - 12\tabcolsep) * \real{0.1026}}
  >{\raggedleft\arraybackslash}p{(\linewidth - 12\tabcolsep) * \real{0.1026}}
  >{\raggedleft\arraybackslash}p{(\linewidth - 12\tabcolsep) * \real{0.1026}}
  >{\raggedleft\arraybackslash}p{(\linewidth - 12\tabcolsep) * \real{0.1026}}
  >{\raggedleft\arraybackslash}p{(\linewidth - 12\tabcolsep) * \real{0.1026}}@{}}
\caption{Median kept holdout AUC under the policy of Table 6 within each
Study 2 pairing, computed exactly over the observed runs. An oracle
choosing on the holdout would raise the mean kept score by at most
0.00013 in any pairing.}\tabularnewline
\toprule\noalign{}
\begin{minipage}[b]{\linewidth}\raggedright
Pairing
\end{minipage} & \begin{minipage}[b]{\linewidth}\raggedleft
1 attempt
\end{minipage} & \begin{minipage}[b]{\linewidth}\raggedleft
3
\end{minipage} & \begin{minipage}[b]{\linewidth}\raggedleft
5
\end{minipage} & \begin{minipage}[b]{\linewidth}\raggedleft
10
\end{minipage} & \begin{minipage}[b]{\linewidth}\raggedleft
Gain, 1 to 3
\end{minipage} & \begin{minipage}[b]{\linewidth}\raggedleft
Gain, 1 to 10
\end{minipage} \\
\midrule\noalign{}
\endfirsthead
\toprule\noalign{}
\begin{minipage}[b]{\linewidth}\raggedright
Pairing
\end{minipage} & \begin{minipage}[b]{\linewidth}\raggedleft
1 attempt
\end{minipage} & \begin{minipage}[b]{\linewidth}\raggedleft
3
\end{minipage} & \begin{minipage}[b]{\linewidth}\raggedleft
5
\end{minipage} & \begin{minipage}[b]{\linewidth}\raggedleft
10
\end{minipage} & \begin{minipage}[b]{\linewidth}\raggedleft
Gain, 1 to 3
\end{minipage} & \begin{minipage}[b]{\linewidth}\raggedleft
Gain, 1 to 10
\end{minipage} \\
\midrule\noalign{}
\endhead
\bottomrule\noalign{}
\endlastfoot
pi, GLM-5.3 Flash & 0.7359 & 0.7460 & 0.7493 & 0.7529 & +0.0101 &
+0.0170 \\
Hermes, GLM-5.3 Flash & 0.7404 & 0.7461 & 0.7499 & 0.7543 & +0.0057 &
+0.0139 \\
OpenCode, GLM-5.3 Flash & 0.7402 & 0.7486 & 0.7518 & 0.7532 & +0.0084 &
+0.0130 \\
OpenCode, DeepSeek 4.1 Flash & 0.7418 & 0.7488 & 0.7506 & 0.7533 &
+0.0070 & +0.0115 \\
Hermes, DeepSeek 4.1 Flash & 0.7437 & 0.7535 & 0.7550 & 0.7607 & +0.0098
& +0.0170 \\
pi, DeepSeek 4.1 Flash & 0.7462 & 0.7567 & 0.7590 & 0.7613 & +0.0105 &
+0.0151 \\
\end{longtable}

\section{Money, tokens and audits}\label{money-tokens-and-audits}

\textbf{Repricing the ledger.} The 312 Study 2 runs read 1.86 billion
input tokens and wrote 25.1 million, a ratio of about 74 to 1, because
the agent re-reads a growing transcript at every step. Table 18 reprices
that ledger at other models' prices. It is a sensitivity calculation on
one transcript: another model would produce a different transcript
length, completion rate and caching profile, so the table does not say
what those models would cost to do the task.

\Needspace{11\baselineskip}

\begin{longtable}[]{@{}
  >{\raggedright\arraybackslash}p{(\linewidth - 6\tabcolsep) * \real{0.4714}}
  >{\raggedright\arraybackslash}p{(\linewidth - 6\tabcolsep) * \real{0.2714}}
  >{\raggedleft\arraybackslash}p{(\linewidth - 6\tabcolsep) * \real{0.1286}}
  >{\raggedleft\arraybackslash}p{(\linewidth - 6\tabcolsep) * \real{0.1286}}@{}}
\caption{Hypothetical repricing of Study 2's token ledger, reasoning
tokens included and billed at the output rate.}\tabularnewline
\toprule\noalign{}
\begin{minipage}[b]{\linewidth}\raggedright
Model
\end{minipage} & \begin{minipage}[b]{\linewidth}\raggedright
Price per M tokens, in / out
\end{minipage} & \begin{minipage}[b]{\linewidth}\raggedleft
Repriced ledger
\end{minipage} & \begin{minipage}[b]{\linewidth}\raggedleft
If 90 percent of input were cached
\end{minipage} \\
\midrule\noalign{}
\endfirsthead
\toprule\noalign{}
\begin{minipage}[b]{\linewidth}\raggedright
Model
\end{minipage} & \begin{minipage}[b]{\linewidth}\raggedright
Price per M tokens, in / out
\end{minipage} & \begin{minipage}[b]{\linewidth}\raggedleft
Repriced ledger
\end{minipage} & \begin{minipage}[b]{\linewidth}\raggedleft
If 90 percent of input were cached
\end{minipage} \\
\midrule\noalign{}
\endhead
\bottomrule\noalign{}
\endlastfoot
GLM-5.3 Flash and DeepSeek 4.1 Flash, as run & \$0.10 / \$0.33 and
\$0.15 / \$0.60 & \$244 & \$54 \\
GLM-5.3 & \$1.40 / \$4.40 & \$2,718 & \$807 \\
Gemini 3.1 Pro & \$2.00 / \$12.00 & \$4,026 & \$1,009 \\
Kimi K3 & \$2.65 / \$13.28 & \$5,268 & \$1,335 \\
Claude Opus 5 & \$5.00 / \$25.00 & \$9,939 & \$2,397 \\
GPT-5.5 Pro & \$30.00 / \$180.00 & \$60,385 & N/A \\
\end{longtable}

\textbf{Cached-input prices} used in the last column, per million
tokens: GLM-5.3 Flash \$0.02, DeepSeek 4.1 Flash \$0.003, GLM-5.3
\$0.26, Gemini 3.1 Pro \$0.20, Kimi K3 \$0.303, Claude Opus 5 \$0.50.
GPT-5.5 Pro lists no cached-input rate, and an unlisted price is not a
zero price, so its entry is left empty.

\textbf{The 23-fold figure} is the ratio of Claude Code's three-run mean
cost to pi's on DeepSeek V4 Flash: 22.5 from unrounded costs, 23 when
rounded. Two of Claude Code's three costs come from its final usage
report and one, for a stalled run, from its request log. That log
repeats each message whenever the conversation is saved again, so
entries must be deduplicated by message id, and its cache-creation field
reads zero on every request.

\textbf{Audit results.} In Study 1's delivered files, the
evaluation-label trace finds the one run that concatenated evaluation
rows into training, and the frame-statistics trace the one whose count
features were computed on the scored data; three runs used the
evaluation set for early stopping, as permitted. Study 2's 312 files
hold five runs that trained on evaluation labels and five that computed
batch features; Study 3's 156 hold three and five. No run did both.

\section{Reproducibility and
disclosure}\label{reproducibility-and-disclosure}

The data repository,
\href{https://github.com/earino/identical-runs-different-results}{github.com/earino/identical-runs-different-results},
holds the task rules and prompts, agent versions and configurations,
model and endpoint identifiers, one row per run for all three studies,
and the holdout. For Studies 2 and 3 it adds every version of the code
each agent delivered, the audit records with each verdict, and the
scoring, statistics and figure code; a run ledger asserts that every
count in the paper reconciles, and VERIFY.md maps each headline claim to
the command that reproduces it. The printed output of every analysis the
paper quotes is committed under \texttt{analysis/results/}, one file per
table, and the README maps each table and figure to its script and file;
retraining a delivered program reproduces its recorded score under the
pinned library versions in \texttt{requirements.txt}. The full run
trees, 43 GB of logs and container state, are not included. The flight
data are public (\citeproc{ref-dataexpo2009}{\emph{Data Expo 2009}
2008}); the repository holds our code, the task data and the agents'
delivered outputs, and redistributes no agent software or model weights.

Claude, working in Claude Code, helped build the benchmark, operate the
runs and draft this paper. It also read the code the compliance traces
flagged and drafted each verdict, which is published with its reason.
The authors chose the studies and the analyses and are responsible for
the verdicts and the text. Claude Code is also one of the six agents
tested in Study 1. Every agent was scored the same way, on the same
hidden holdout.

\end{document}